\documentclass[journal,svgnames]{vgtc}                     

\ifpdf
  \pdfoutput=1\relax                   
	\usepackage{dblfloatfix}    
  \usepackage{graphicx}                
  \DeclareGraphicsExtensions{.pdf,.png,.jpg,.jpeg} 
\else
  \usepackage{graphicx}                
  \DeclareGraphicsExtensions{.eps}     
\fi%

\usepackage{xspace}

\usepackage[normalem]{ulem}            
\newcommand{\acronym}{DiffFit\xspace}

\onlineid{0}

\vgtccategory{Research}

\vgtcpapertype{please specify}

\newcommand{\mytitle}{\acronym: Visually-Guided Differentiable Fitting\\ of Molecule Structures to a Cryo-EM Map}
\title{\mytitle}

\author{%
  \begin{picture}(0, 0)(0, 0)%
  \put(0, 100){\small This is an updated version of the original paper (DOI: \href{https://doi.org/10.1109/TVCG.2024.3456404}{10.1109/TVCG.2024.3456404}, published in the IEEE Transactions on Visualization and Computer}%
  \put(0, 92){\small Graphics 31(1)) that includes the changes from the subsequent errata (DOI: \href{https://doi.org/10.1109/TVCG.2024.3502911}{10.1109/TVCG.2024.3502911}, published in volume 31, number 2).}
  \end{picture}\authororcid{Deng Luo }{0000-0003-4610-8730},
  \authororcid{Zainab Alsuwaykit}{0009-0009-7444-445X}, 
  \authororcid{Dawar Khan}{0000-0001-5864-1888}, 
  \authororcid{Ondřej Strnad}{0000-0002-8077-4692}, 
  \authororcid{Tobias Isenberg}{0000-0001-7953-8644}, and 
  \authororcid{Ivan Viola}{0000-0003-4248-6574} 
}

\authorfooter{
  \item
  	Deng Luo (\raisebox{-.5pt}{\includegraphics[height=6pt]{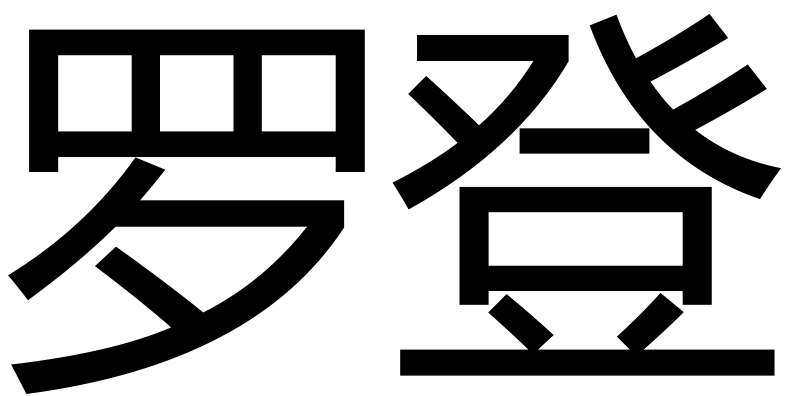}}), Zainab Alsuwaykit (\raisebox{-.5pt}{\includegraphics[height=6pt]{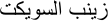}}), Dawar Khan (\raisebox{-.5pt}{\includegraphics[height=6pt]{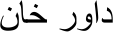}}), Ondřej Strnad, and Ivan Viola are with the Visual Computing Center at King Abdullah University of Science and Technology (KAUST), Saudi Arabia.
  	E-mail: \{deng.luo\discretionary{\,\textbar}{}{\,\textbar\,}zainab.alsuwaykit\discretionary{\,\textbar}{}{\,\textbar\,}dawar.khan\discretionary{\,\textbar}{}{\,\textbar\,}ondrej.strnad\discretionary{\,\textbar}{}{\,\textbar\,}ivan.viola\}@kaust.edu.sa.
\item Tobias Isenberg is with Université Paris-Saclay, CNRS, Inria, LISN, France. E-mail: given\_name.family\_name@inria.fr.
}

\abstract{%
We introduce \acronym, a differentiable algorithm for fitting protein atomistic structures into an experimental reconstructed \ac{Cryo-em} volume map. In structural biology, this process is necessary to semi-automatically composite large mesoscale models of complex protein assemblies and complete cellular structures that are based on measured \acs{Cryo-em} data. 
The current approaches require manual fitting in three dimensions to start, resulting in approximately aligned structures followed by an automated fine-tuning of the alignment. The \acronym approach enables domain scientists to fit new structures automatically and visualize the results for inspection and interactive revision. The fitting begins with differentiable three-dimensional (3D) rigid transformations of the protein atom coordinates followed by sampling the density values at the atom coordinates from the target \acs{Cryo-em} volume. To ensure a meaningful correlation between the sampled densities and the protein structure, we proposed a novel loss function based on a multi-resolution volume-array approach and the exploitation of the negative space. This loss function serves as a critical metric for assessing the fitting quality, ensuring the fitting accuracy and an improved visualization of the results. We assessed the placement quality of \acronym with several large, realistic datasets and found it to be superior to that of previous methods. We further evaluated our method in two use cases: automating the integration of known composite structures into larger protein complexes and facilitating the fitting of predicted protein domains into volume densities to aid researchers in identifying unknown proteins. We implemented our algorithm as an open-source plugin (\href{https://github.com/nanovis/DiffFit}{\texttt{github\discretionary{}{.}{.}com\discretionary{/}{}{/}nanovis\discretionary{/}{}{/}DiffFit}}) in ChimeraX, a leading visualization software in the field. All supplemental materials are available at \href{https://osf.io/5tx4q/}{\texttt{osf.io/5tx4q}}.

}

\keywords{Scalar field data, algorithms, application-motivated visualization, process/workflow design, life sciences, health, medicine, biology, structural biology, bioinformatics, genomics, cryo-EM.}

\teaser{
  \centering
  \includegraphics[width=\linewidth]{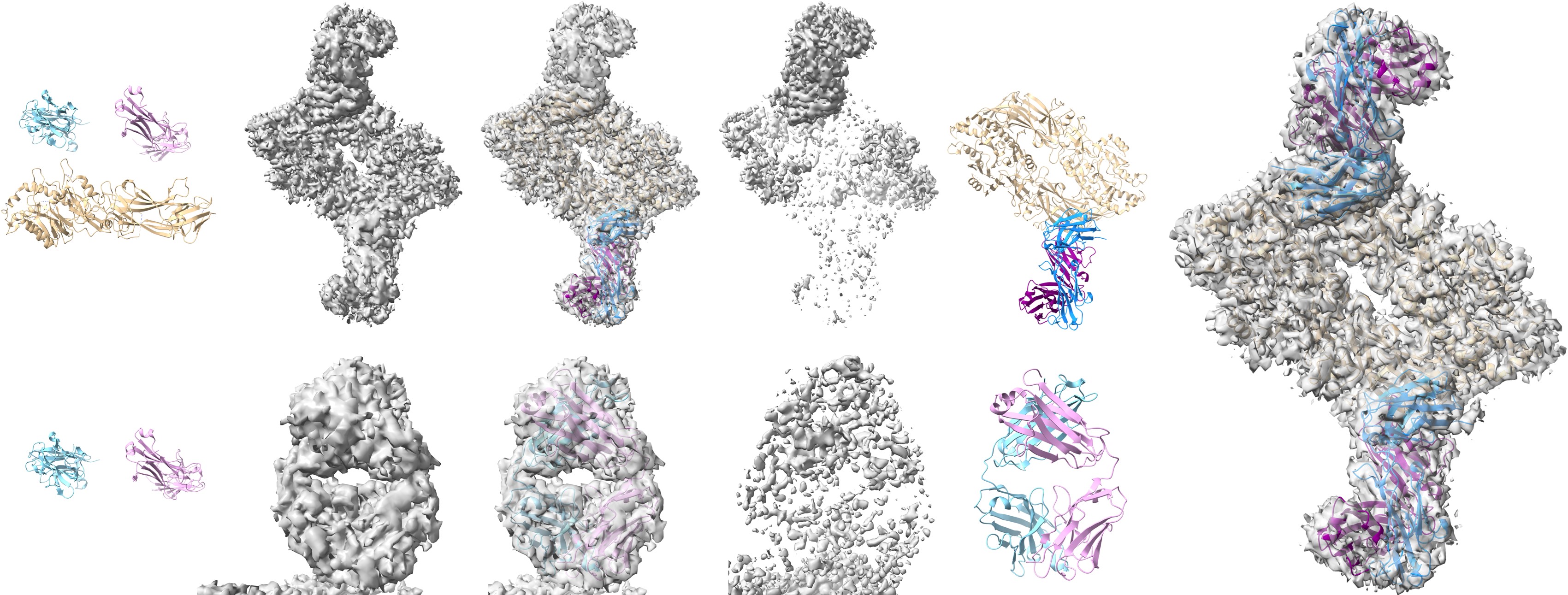}
  \caption{%
  	Compositing a protein (\href{https://doi.org/10.2210/pdb8SMK/pdb}{PDB 8SMK} \cite{Zhou:2024:ADI}) from its three unique chains. Top row from left to right: three input chains, input target volume, the best fits in the first fitting round, the remaining voxels after \emph{zeroing-out}, and the fitted chains. Bottom row from left to right: two remaining input chains, remaining region of interest in the target volume from the first round, the best fits in the second round, the remaining voxels after \emph{zeroing-out}, the fitted chains. Right: the final composited structure overlaid on the original target volume \change{(RMSD: 0.138)}. The involved computation takes \change{10} seconds in total, and the human-in-the-loop interaction takes $\approx$\,\change{3} minutes.%
  }
  \label{fig:teaser}
}

\graphicspath{{figs/}{figures/}{pictures/}{images/}{./}} 

\usepackage{tabu}                      
\usepackage{booktabs}                  
\usepackage{lipsum}                    
\usepackage{mwe}                       

\usepackage{mathptmx}                  
\usepackage{cuted}

\usepackage{acro}
\usepackage{algorithm}
\usepackage{algorithmicx}
\usepackage[algo2e]{algorithm2e}
\usepackage{algcompatible}
\usepackage{wrapfig}
\usepackage{textcomp}
\usepackage{xspace}
\usepackage{ccicons}
\usepackage{amsmath}
\usepackage{setspace}
\let\Algorithm\algorithm
\renewcommand\algorithm[1][]{\Algorithm[#1]\setstretch{1.5}}
\usepackage{placeins} 
\usepackage{enumitem}
\usepackage{balance}
\usepackage{lineno}
\usepackage{subfigure}
\usepackage{multirow}
\newcommand{\etal}{\textit{et al.}\xspace}
\newcommand{\eg}{e.\,g.}
\newcommand{\ie}{i.\,e.}

\newcommand{\change}[1]{\textcolor{FireBrick}{#1}}
\renewcommand{\change}[1]{#1}

\usepackage{pdfpages}

\DeclareAcronym{cvt}{
  short=CVT,
  long=Centroidal {V}oronoi tessellation,
}
\DeclareAcronym{Cryo-em}{
  short=cryo-EM,
  long=Cryo-Electron Microscopy ,
} 
\DeclareAcronym{Cryo-et}{
  short=cryo-ET,
  long=Cryo-Electron Tomography,
}  

\DeclareAcronym{pca}{
  short=PCA,
  long=principal component analysis,
}

\begin{document}


\firstsection{Introduction}

\maketitle

As humans we have been striving for centuries or even millennia to understand, as \href{https://en.wikipedia.org/wiki/Faust}{Faust} stated, ``\textit{was die Welt im Innersten zusammenh{\"a}lt} [what binds the world, and guides its course]'' \cite{Goethe:1808:F}. Among other directions of scientific inquiry, this quest applies to the inner workings of the biological world, particularly to how biological processes at tiny scales work and how they keep us alive. In the field of structural biology, researchers have traditionally relied on techniques such as X-ray crystallography or nuclear magnetic resonance spectroscopy to understand the actual molecular composition of cells and organelles---yet with the limitation that these could only provide (still impressive and highly useful) estimates or manually constructed models of the structure of actual biological samples (e.g., \cite{Schaefer:2024:InVADo,Ulbrich:2023:sMolBoxes,Kutak:2022:Vivern,Skanberg:2016:RMV,Nguyen:2012:Modeling,Duran:2019:VLM}). The recent \ac{Cryo-em} approach \cite{Nakane2020SingleparticleCA}, however, enables researchers to visualize biomolecules in \emph{actual samples} at near-atomic resolution. In addition, over decades, the Protein Data Bank (PDB) initiative has collected thousands of molecular models of the building blocks of cells or organelles studied in structural biology. Researchers are thus on the brink of assembling the molecular composition of actual samples at the ground-truth level.

To achieve this type of assembly, researchers do not only need to interactively visualize molecular data, for which tools \cite{Skanberg:2023:VIAMD} exist, but also to faithfully place 3D models of known molecular building blocks, such as those from PDB data and the AlphaFold predicted library \cite{Jumper:2021:HAP}, into the captured \ac{Cryo-em} datasets.
Thus far, the fitting process involves a substantial time commitment and numerous manual interventions by domain experts, rendering this process ineffective. The complexity and size of the involved molecules, combined with the variability and noise inherent in \ac{Cryo-em} data, pose substantial obstacles. In contrast, a fully automatic process is also not ideal because the existence of local minima (wrong placement of compositing proteins) requires domain experts to verify each placement using their knowledge and experience. Fully automated methods are currently far from feasible. Instead, an optimal balance between user interaction and automation is required.

For this purpose, we developed \acronym, an automated differentiable fitting algorithm coupled with visual inspection and decision-making, designed to optimize alignment between protein structures and experimental reconstructions of volumes (\ie, \ac{Cryo-em} maps). Our technique works in one-to-one and many-to-one fitting scenarios, in which multiple protein subunit structures are precisely aligned with a single, large, experimentally reconstructed volume. The \acronym method is iterative and gradually introduces the source protein structures into the target volumes to assemble the composition of the molecular subunits step by step. By employing advanced strategies such as volume filtering, multiresolution volumes, and negative space utilization we constructed a loss function to quantify the fitting accuracy during the iterations and for the final decision-making. 
This loss function helps us to iteratively reduce the differences between the two representations---volumetric and atomistic---until we achieve the desired fit. 
Our visually-guided fitting procedure eliminates the need for domain experts to manually place structures as they assemble the protein structures into the \ac{Cryo-em} map. It thus significantly accelerates the process into a manageable interactive procedure, delivering precise results for visualizing and analyzing complex, real-world protein structures, ultimately facilitating large-scale structural modeling initiatives.
In summary, we contribute:
\begin{itemize}[noitemsep]
\item a differentiable fitting algorithm designed to fit multiple molecular subunits to a single reconstructed \ac{Cryo-em} volume;
\item a human-in-the-loop strategy providing visual inspection and decision-making in an iterative structure assembly cycle;
\item a novel loss function and data processing that calculates new updates in each iteration to expedite algorithm convergence and quantify the fitting accuracy; and
\item three use-case scenarios of fitting one or multiple known subunits or identifying yet unknown subunits as part of the molecular assembly.
\end{itemize}

\section{Background and related work}
\label{sec:related-work}



We rely on \ac{Cryo-em} data, so below we first briefly provide essential relevant background. We then describe past work on image registration and model fitting and show how both relate to our own research. 

\subsection{Brief background}

Structural biology employs various techniques to understand how atoms are arranged in macromolecular complexes, ranging from 60\,kDa (\ie, 4,472 atoms; \href{https://doi.org/10.2210/pdb6NBD/pdb}{PDB 6NBD} \cite{Herzik:2019:HSD}) to 50,000\,kDa (3,163,608 atoms; \href{https://doi.org/10.2210/pdb8J07/pdb}{PDB 8J07} \cite{Walton:2023:ASR}). These techniques are essential for the study of processes in living cells---\ac{Cryo-em} being a particularly powerful one~\cite{Bai2015,MALHOTRA2019Modeling,Lasker2007EMatch}. With \ac{Cryo-em}, bioscientists can capture images of flash-frozen biological specimens using an electron microscope, preserving their natural structure without the interference of staining or fixing \cite{Chen2023}, which would otherwise interfere with the sample. 
These images are then used to construct \ac{Cryo-em} 3D volumes or maps using the \emph{single particle method}, which aligns thousands of projections from structurally identical molecular instances into a single map using the Fourier slice-projection theorem. This map represents the electron density of the sample, which can be used to infer the atom positions within the molecule. 

Subsequently, the bioscientists need to build accurate atomistic or molecular models that match the electron density map obtained from the \ac{Cryo-em} process to gain insight into molecular function and interactions. This process involves mapping or fitting known sub-molecules into their corresponding positions within the map. The objective is to achieve an optimal correspondence between the model and the experimental or simulated volume, revealing the organization of molecules in 3D space, including single molecules, complexes, and the placement of small molecules and ligands into binding sites. Molecular models are available in the Protein Data Bank (PDB, \href{https://www.rcsb.org/}{\texttt{rcsb\discretionary{}{.}{.}org}}), accessible in various formats such as PDB, Crystallographic Information File (CIF), and mmCIF (macromolecular CIF). As of now, the fitting is typically achieved through manual placement, alignment, and comparison with the density maps. The manual nature of this process makes it time-con\-su\-ming and tedious, and can only be performed by expert biologists. To address this challenge, numerous approaches have been developed to automate the fitting process, which largely focus on image registration as the foundation and explore methods to streamline 3D model construction, as we review next.

\subsection{Image registration and geometric fitting}

The fitting of 3D structures into captured or simulated volumes relates to the problem of image registration in image processing.  It entails aligning two images, originating from the same or from different modalities, within a shared reference frame \cite{Hill2001, Fu2020}. This process involves feature extraction, determining transformations, and assessing accuracy through metrics. Scale-invariant features from images~\cite{Lowe2004}, for example, can facilitate matching across a diverse set of views, despite significant distortions or variations. This process involves detecting invariant keypoints using the difference-of-Gaussian function, determining locations and scales, assigning directions based on local gradients, and measuring gradients within selected scales around each keypoint. Extracted features are stored in a database, to make it possible for them to be matched with new images using fast nearest-neighbor algorithms, with applications including object recognition.

Among the many applications of the process, physicians rely on various imaging modalities to diagnose patients, each capturing images with differing orientation. Image registration addresses this variability by aligning images within a unified frame by optimizing parameters like orientation and translation. 
Medical image registration is an active research area which encompasses diverse methods, including techniques based on cross-cor\-re\-la\-tion \cite{Didon1995,Malinsky2013} and mutual information \cite{Pluim2003,Mattes2003,Thevenaz2000,Klein2009}. 
Shang \etal~\cite{SHANG2006}, for example, introduced a method for medical image registration using \ac{pca} neural networks to extract feature images and compute rotation angles and translation parameters by aligning the first principal directions and centroids in a simple and efficient way. For complex spatial transformations, another recent approach~\cite{arora2024integration} uses Kernel \ac{pca} and Teaching-Learning-based optimization (TLBO) for multi-modal image registration.  
In our case, similar to these methods, transformations and alignments have to be determined to fit the atomistic model into a volumetric map. We can thus also use optimization techniques in \ac{Cryo-em} map fitting to refine the fit and optimize parameters such as orientation and translation---which we demonstrate in our work. The major difference to image registration is that, in our workflow, we fit two different data representations, where one is a sub-part of the whole that is potentially present at multiple locations in the target volume.

Model-to-data fitting, which is necessary for \ac{Cryo-em} data, has also been investigated in depth in computer graphics and pattern recognition~\cite{Reddy2020Differentiable,K2018Direct}, with applications in architectural geometry, virtual and augmented reality, robotics, and various other fields \cite{Falk2023,Agnes2018Spline,kou2024adaptive,Yeh2011Template}---in addition to structural biology. The key challenges in geometric fitting include accuracy, efficiency, robustness, and usability of the fitting module~\cite{Agnes2018Spline,kou2024adaptive}. Structural biology, in contrast, has special challenges such as noisy data, non-geometric shapes, and large data sizes so that geometric fitting methods are not directly applicable.

Yet our \acronym algorithm still relates to techniques from computer graphics and pattern analysis. The differentiable compositing technique proposed by Reddy \etal \cite{Reddy2020Differentiable}, in particular, offers valuable insights into addressing fitting challenges as well as manipulating and understanding image patterns. With \emph{differentiable compositing} we can handle patterns effectively, outperforming state-of-the-art alternatives in pattern manipulation \cite{Zhou2018Nonstationary,Shaham2019SinGANICCV}. Reddy \etal's method~\cite{Reddy2020Differentiable} discovers complex patterns by aligning elements with their own position and rotation, and facilitates refinement based on similarity to the target for precise adjustment. In addition, Reddy \etal use a multi-resolution pyramid---relevant for handling the multi-resolution volumetric data in our domain. Their method~\cite{Reddy2020Differentiable}, however, is restricted to certain pattern types, requires manual element marking, and may not always find the best solution, leading to orientation errors and missed elements. Nevertheless, we built our solution on top of their differentiable compositing.

Another approach, spline surface fitting \cite{Agnes2018Spline}, enhances the smoothness in aircraft engine geometry reconstruction by concurrently approximating point and normal data, ensuring boundary smoothness and optimal convergence, while exploring the effects of norm-like functions on error measurement. A further recently proposed adaptive spline surface fitting method~\cite{kou2024adaptive}, supported by empirical evidence, employs surface meshes for high-precision CAD applications. The reliance on control meshes of this approach, however, limits its applicability to irregular topologies and compromise the preservation of sharp features. All these methods have common objectives and tasks such as similarity measures, pattern matching, fitting, and geometric transformations; they thus can serve as a motivation and starting point toward our goals in structural biology. Structural biology data, however, often consists of large, complex structures without regular shapes such as CAD models or easy representations in geometric meshes with smooth surfaces so the aforementioned methods are not directly applicable to our data.
 
\subsection{Fitting in structural biology}

Existing fitting methods for structural biology can broadly be categorized into manual, semi-automated, and automated approaches, each with its own advantages and challenges when used for aligning molecular models with \ac{Cryo-em} density maps. Manual or semi-automated methods naturally involve human intervention, yet they provide control and precision---which is particularly beneficial in the structural analysis of complex datasets or when specific adjustments are needed for accuracy. For example, \textit{UCSF ChimeraX}~\cite{pettersen2004ucsf}, a popular tool for molecular manipulation and visualization, includes the \emph{fitmap} technique \cite{Thomas2007}. It suggests multiple possible placements of the atomistic model on the density map and then asks the user to make the final decision. The fitting process alternates between rigid-body rotation and gradient descent translation, maximizing the alignment between the atomic model and the density data by optimizing the sum of density values. Similarly, MarkovFit~\cite{Alnabati2022} is another technique that is often used for placing atomic-level protein structures within \ac{Cryo-em} maps of moderate to low resolutions. It uses fast Fourier transform (FFT) for the conformational exploration and Markov random fields (MRF) for efficient representation of subunit interactions. Its use of Markov random fields also facilitates the probabilistic assessment of the fitted models.
Nonetheless, all of these manual or semi-automatic approaches are time-consuming and require a significant level of expertise.

To tackle this challenge and to automate the fitting process, researchers have developed methods that rely on deep learning (DL) \cite{Xu2019,Terwilliger2018,Wang2020}. $A^2$-Net by Xu \etal \cite{Xu2019}, for example, uses DL to accurately determine amino acids within a 3D \ac{Cryo-em} density volume. It employs a sequence-guided Monte Carlo Tree Search (MCTS) to traverse candidate amino acids, considering the sequential nature of amino acids in a protein. The authors divide the problem of molecular structure determination into three subproblems: amino acid detection in the density volumes, assignment of atomic coordinates to determine the position of each amino acid, and main chain threading to resolve the sequential order of amino acids that form each protein chain. A remarkable speed improvement was also demonstrated by Xu \etal, being 100\texttimes{} faster to find solutions at runtime than existing methods \cite{frenz2017rosettaes,wang2015novo}, and achieving a high accuracy of $89.8\%$. In addition, they introduced the $A^2$ dataset with 250,000 amino acids in 1,713 \ac{Cryo-em} density volumes, with a resolution of 3 $A^{\circ}$, pioneering automated molecular structure determination training benchmarks.


Another recent method by Mallet \etal, CrAI, uses machine learning (ML) to find antibodies in cryo-EM densities \cite{Mallet2023}. The authors formulate the objective as an object detection problem, using the structural properties of Fabs (Fragment antigen-binding) and VHHs (single-domain antibodies). Furthermore, DeepTracer \cite{Pfab2021} is a fully automated DL-based method designed to determine the all-atom structure of a protein complex using its high-resolution \ac{Cryo-em} map and amino acid sequence. This method employs a customized deep convolutional neural network primarily for the precise prediction of  protein structure, including the locations of amino acids, backbone, secondary structure positions, and amino acid types. The reprocessed \ac{Cryo-em} maps are the input to the neural network, which transforms the output into a protein structure. Despite yielding accurate outcomes, the resulting atomistic structures may exhibit geometric issues, local fit-to-map discrepancies, misplaced side chains, or errors in tracing and\discretionary{/}{}{/}or connectivity. All DL-based techniques require a substantial amount of time for training (as opposed to their runtime performance) and rely on large training datasets of \ac{Cryo-em} volumes and manually fitted sub-mo\-le\-cules---which is why we do not resort to DL approaches.
%
%

An alternative to DL is map-to-map alignment, which is used to accurately align 2D or 3D maps to facilitate comparison and analysis of spatial structures or features within the maps. CryoAlign \cite{he2024} is a \ac{Cryo-em} density map alignment method that achieves a fast, accurate, and robust comparison of two density maps based on local spatial feature descriptors. This approach involves sampling the density map to generate a point cloud representation and extracting key points by clustering based on local properties. CryoAlign then calculates local feature descriptors to capture structural characteristics, reducing the number of points considered and improving efficiency. By employing a mutual feature-matching strategy, CryoAlign establishes correspondences between keypoints in different maps and uses iterative refinement to enhance alignment. A combination of fast rotational matching search based on spherical harmonics and translational scans \cite{garzon2006} yields accurate fitting results in seconds or up to a few minutes. This ADP\_EM approach is particularly reliable in fitting X-ray crystal structures to low-resolution density maps, with reduced docking times and while maintaining a thorough 6D exploration with fine rotational sampling steps to find valid docking solutions.

%

In our work, we design a differentiable optimization method for fitting atomistic structures into volumetric data, with the goal of precise fitting with fast-enough computation to be applicable in semi-automatic fitting in the standard tool ChimeraX. For this purpose, we make use of the PyTorch capabilities for GPU parallel computing, trilinear interpolation sampling in volumetric data, and auto differentiation.

\section{Method}
\label{sec:method}
\begin{figure*}[!t]
  \centering
  \includegraphics[width=\linewidth]{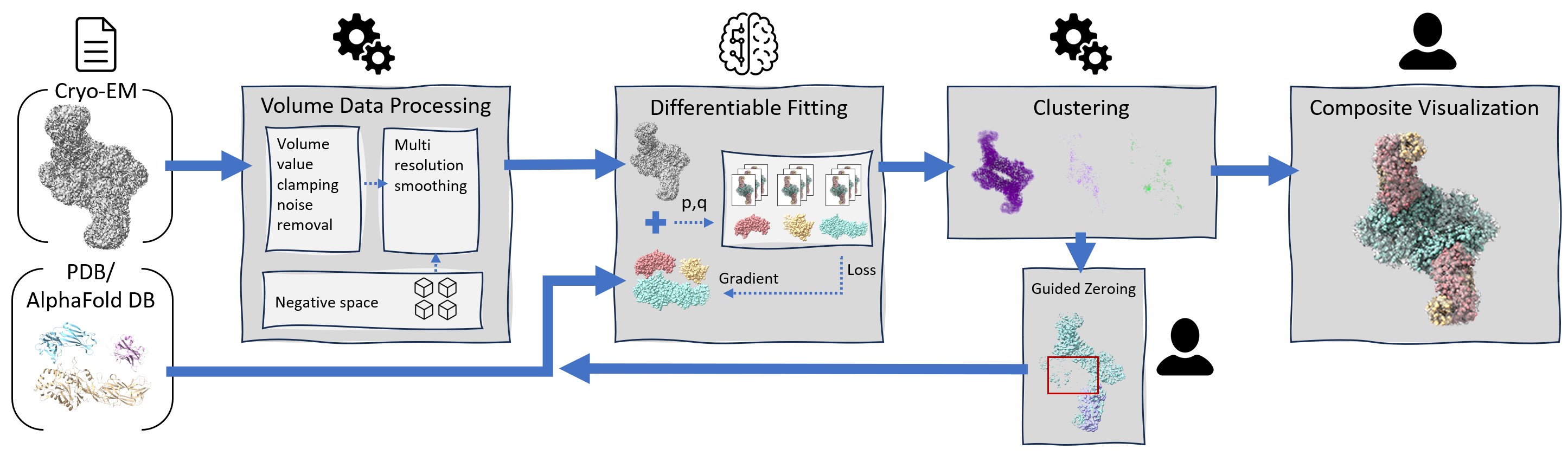}
  \caption{\acronym workflow. The target \ac{Cryo-em} volume and the structures to be fit on the very left serve as inputs, which are passed into the novel volume processing, followed by the differentiable fitting algorithm. The fitting results are then clustered and inspected by the expert. The expert may zero out voxels corresponding to the placed structures and feed the map back iteratively as input for a new fitting, round until the compositing is done.}
  \label{fig:method_overview}
\end{figure*}
We begin to describe our approach by explaining our process of differentiable structure fitting, before we show how it can be used for visually-guided fitting. After discussing these conceptual aspects, we also briefly discuss implementation details.

\subsection{Differentiable structure fitting}
Given a \ac{Cryo-em} map, the domain practitioners---bioscientists---do not know the precise location and orientation parameters that govern where and how a protein's sub-structures fit together. For some regions in the map, the bioscientists may not even know which protein subunits are supposed to be present. Our goal is to develop a new approach that addresses both of these domain tasks. For certain protein subunits, bioscientist are highly confident about their presence in the map. In such cases, our technique will aid in the identification of their respective placement parameters. Second, for the regions with unknown protein subunits, the task of our technique is to identify potential protein subunit candidates from a large database that best fit the \ac{Cryo-em} map region. 

\subsubsection{Inspiration and approach overview}

We base our approach on the previously mentioned 2D differentiable compositing approach by Reddy \etal \cite{Reddy2020Differentiable}, which discovers pattern structure from wallpaper-like textures containing repetitive patterns made out of elementary patches. We first review their approach, before we describe how we build up our solution on top of their technique. In their case, given a 2D image, which is a composite of multiple small element patches, the task is to identify the number of occurrences of each patch and the placement parameters for each existing patch. The parameters include the type of patch out of several known patches as well as position, orientation, and depth. Their solution distributes tens to hundreds of patches in the image and uses the differentiable optimization methodology to translate and reorient patters such that they correspond to the appearance of the patterns in the input wallpaper image. Each single instance $E_i$ (out of total number of $\omega$ instances) of a pattern is stored in a layer $J_i$ for each patch instance by sampling from that patch, with the translation, rotation, and patch-pattern type probability taken into account:\vspace{-1ex}
%

\begin{equation}
J_i(\mathbf{x})=f_t\left(\mathbf{x}, E_i\right)=\sum_{j=1}^o \frac{e^{t_i^j}}{\sum_{k=1}^o e^{t_i^k}}  h_j\left(R_{\theta_i}^{-1}\left(\mathbf{x}-\mathbf{c}_i\right)\right)
\end{equation}\vspace{-1ex}

\newlength{\lineskiplimitbackup}%
\setlength{\lineskiplimitbackup}{\lineskiplimit}%
\setlength{\lineskiplimit}{-\maxdimen}%
\noindent where $f_t\left(\mathbf{x},E_i\right)$ is a differentiable function using the expected value over patch-pattern type probabilities stored in a tuple $\mathbf{t}$ representing all $o$ patch patterns; softmax $e^{t_i^j}/\sum_{k=1}^o e^{t_i^k}$ over type logits define the patch-pattern type probabilities; $h_j$ is the image patch sampler function; $\mathbf{x}$ is the image location; $\mathbf{c}_i$ is center location of patch element $E_i$; and $R_{\theta_i}^{-1}$ is the inverse of a 2\,\texttimes\,2 matrix rotation with angle $\theta_i$.
 
\setlength{\lineskiplimit}{\lineskiplimitbackup}%

The solution image results from compositing of all instances together using $f_c$ compositing function so that each known patch-pattern type is present multiple times with various positional parameters. Patches can overlap other patches, which leads to partial or full occlusion of a certain pattern. This is characterized by $v_i(\mathbf{x}), v \in \{0,1\}$, which is the visibility of layer $i$ at image location $\mathbf{x}$:\vspace{-1ex}
\begin{equation}
I(\mathbf{x})=f_c\left(\left\{J_i(\mathbf{x})\right\}_i\right)=\sum_{i=0}^\omega J_i(\mathbf{x}) v_i(\mathbf{x})
\end{equation}
This solution image is compared with the input image in an optimization, where the parameters of all patch instances are updated in every iteration. The so\-lu\-tion image then becomes increasingly similar to the in\-put image. Reddy \etal define the $L^2$ distance loss $L_d$ for the optimizer as:\vspace{-1ex}
\begin{equation}
L_d(A, I)=\frac{1}{P} \sum_{p=1}^P\left\|A\left(\mathbf{x}_p\right)-I\left(\mathbf{x}_p\right)\right\|_2^2
\end{equation}
where the sum is over the number of all pixels $P$ in the image. $A$ is the input image and $I$ is the composited solution image.
The optimal elements $\mathcal{E}^*$ are then found by minimizing  loss $L_d$ over the entire set of elements $\mathcal{E} = \{E_0, .., E_\omega\}$
:\vspace{-1ex}
\begin{equation}
\mathcal{E}^*=\underset{\mathcal{E}}{\arg \min } ~L_d(A, f_c(\mathcal{E})) .
\end{equation}

\begin{figure*}[t]
  \centering
	\setlength{\subfigcapskip}{-2.5ex}
  \subfigure[\hspace{\columnwidth}]{\label{fig:clustering_filtering:a}\includegraphics[height=4cm]{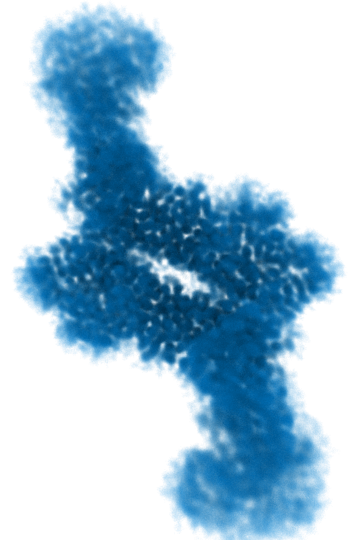}}\hfill
  \subfigure[\hspace{\columnwidth}]{\label{fig:clustering_filtering:b}\includegraphics[height=4cm]{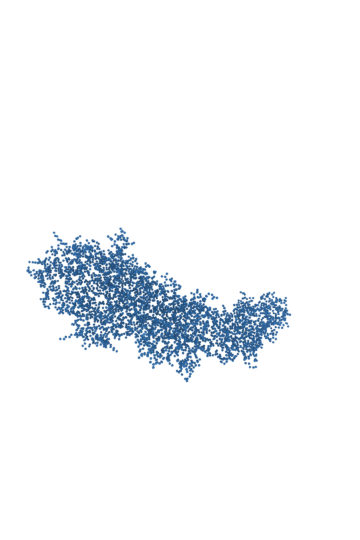}}\hfill
  \subfigure[\hspace{\columnwidth}]{\label{fig:clustering_filtering:c}\includegraphics[height=4cm]{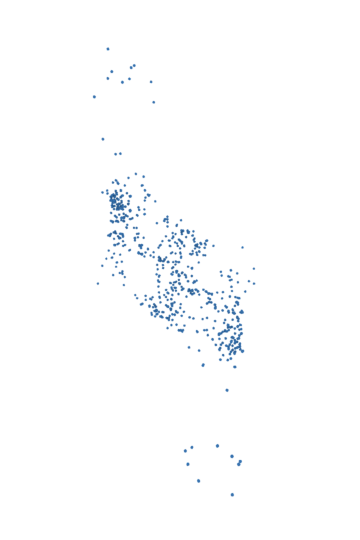}}\hfill
  \subfigure[\hspace{\columnwidth}]{\label{fig:clustering_filtering:d}\includegraphics[height=4cm]{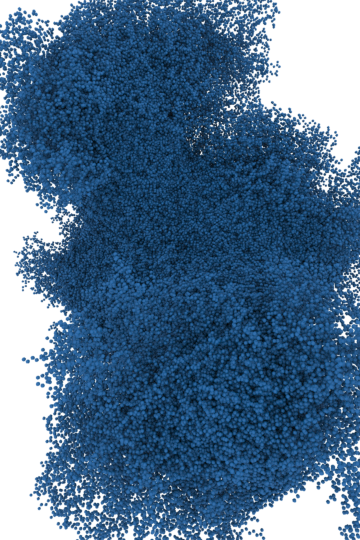}}\hfill
  \subfigure[\hspace{\columnwidth}]{\label{fig:clustering_filtering:e}\includegraphics[height=4cm]{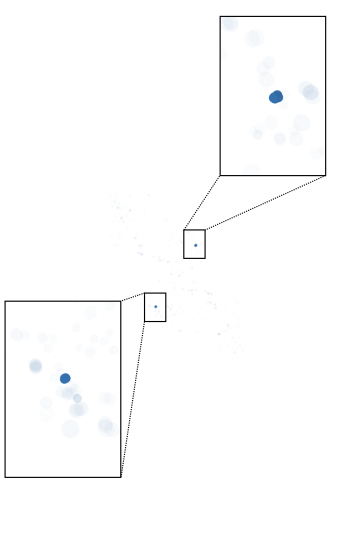}}\hfill
  \subfigure[\hspace{\columnwidth}]{\label{fig:clustering_filtering:f}\includegraphics[height=4cm]{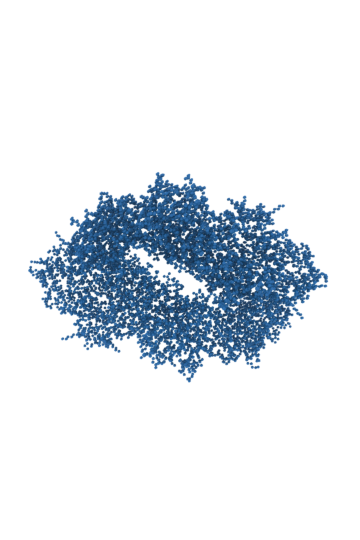}}\vspace{-2ex}
  \caption{Clustering and filtering: (a) target volume, (b) atom coordinates of the source structure, (c) positions of 1000 fit results (the dots are clustered, hiding the number of results), (d) 1000 instances of the structure, (e) positions of 1000 fit results with a transparency level set based on an exponential scaling of the \change{sum of sampled density} metric (two clusters stand out\change{, as in the zoom-in insets}), and (f) instances of the structure at those two clusters.}
  \label{fig:clustering_filtering}
\end{figure*}

Reddy \etal's patch-pattern fitting problem is similar to ours in the sense that in both cases we are compositing element instances into a scene. But Reddy \etal's differentiable compositing approach cannot be directly applied to the structural biology domain to solve the protein fitting problem for the following reasons:
\begin{enumerate}[noitemsep,left=0pt .. \parindent]
    \item the pattern image and element patches are defined in 2D with layers, while the \ac{Cryo-em} map and protein subunits are defined in 3D; 
    \item the pattern image and element patches are of the same representation, i.e., 2D grid data, while the \ac{Cryo-em} map and protein subunits are of different representations---one is a 3D volume while the other is a set of atom coordinates that can be regarded as a point cloud; 
    \item the instance patches in differentiable compositing are all of the same size, while the protein subunits differ in numbers of atoms;
    \item differentiable compositing expects the patches to overlap, while protein subunits do not spatially overlap; and 
    \item forming 1000 layers of 2D images is possible to fit into the current GPU memory while forming 1000 3D volumes is prohibitive with the currently available GPU memory.
\end{enumerate}
Initially, we attempted to align our problem better with differentiable compositing by first \emph{simulating} a \ac{Cryo-em} map from the atomistic point cloud of the protein model and then fitting the simulated map to a target map. That way the representational discrepancy (see reason (2) above) is eliminated. That approach, however, was only successful for trivial cases, while for real-world scenarios it frequently fell into local minima. To illustrate this point, we provide several exemplary volume-only based molecular fitting videos for interested readers in our supplementary material at \href{https://osf.io/5tx4q/}{\texttt{osf.io/5tx4q}}.

Driven by the successful fitting cases from many experiments, we gradually built several novel strategies on top of the differentiable compositing that effectively address the problem of molecular structure fitting. Most notably, we address the substantially higher complexity of our scenario based on the knowledge, experience, and deep insight of the target audience: the bioscientists. Our solution is thus based on a human-in-the-loop strategy and we propose a fast and robust visual analytics approach, \acronym, with two main steps: (1) an automated excessive molecular fitting and (2) a visual inspection and filtering of the fitted results by the bioscientists. Both consecutive steps are building blocks of a visual analytics feedback loop, in which multiple proteins are iteratively composited to fit the underlying \ac{Cryo-em} volume. 

We schematically present our \acronym workflow in \autoref{fig:method_overview}. First, we seed an excess amount of all compositing molecules in the volume scene. If this simultaneous fitting all molecules exceeds the available GPU memory, we sort the molecules by atom count and partition them into batches. Then, we fit the batches of molecules within several iterations in descending atom-count order. This fitting relies on a novel loss function that calculates the average density value from the densities that we sample for each atom. The differentiable property of our fitting scenario allows us to optimize based on gradient-descent. In addition, we associate each fit with a numerical value that characterizes the fit quality. For this purpose, we create a simulated \ac{Cryo-em} map of each fitted molecule and calculate the correlation of the simulated densities using the real \ac{Cryo-em} map densities. Then, we collect the fitting results and cluster them based on positional and orientational parameters. For each cluster, we select one representative fit---the fit with the highest correlation value. Then, we sort the clusters by their representative correlations and interactively visualize them in ChimeraX to allow the bioscientists to inspect the solutions. Once they verify a given molecular placement, we disable molecule placements in the respective regions in the following iterations by setting the  voxels covered by the molecular fit in the \ac{Cryo-em} map volume to zero. We thus gradually erase the successful placements from the map, forcing the following placements to search for a fit in non-zero volume locations. Once the feedback update in the map is completed, we perform the next fitting iteration with another molecular structure. We repeat this workflow pattern until the map has nearly all voxels zeroed out, and the entire sub-unit placement of the complex molecular structure is complete. Below we introduce the details of \acronym, in the following order: (1) sample one coordinate, (2) fit one placement of one molecule, (3) fit multiple placements of one molecule, and (4) fit multiple placements of multiple molecules.

\subsubsection{Sampling of one coordinate}

Because our task is to determine the optimal alignment of an atomistic molecular structure to the reconstructed \ac{Cryo-em} volume map, we determine the optimal fit characterized by two rigid-body transformation parameters: a translational offset $\mathbf{p}$ and a rotation. We represent the rotation by a quaternion $\mathbf{q}$ or its corresponding rotation matrix $M_{\mathbf{q}}$. The position $\mathbf{x}_{i}$ corresponds to the center point of an atom $i$ in the molecular subunit. To calculate the fit, we transform every atom position in one subunit according to the rotation and translational offset:
\begin{displaymath}
T(\mathbf{x}_i) = M_{\mathbf{q}} \cdot \mathbf{x}_i + \mathbf{p}.
\end{displaymath}

We sample a density value $D$ of the atom to be placed at position $T(\mathbf{x}_i)$ from a scalar volume $V$ using trilinear interpolation as follows: 
\begin{displaymath}
D(T(\mathbf{x}_i)) = S (T(\mathbf{x}_i), V).
\end{displaymath}

\begin{figure*}
	\centering
	\includegraphics[width=\textwidth,height=.5\textwidth]{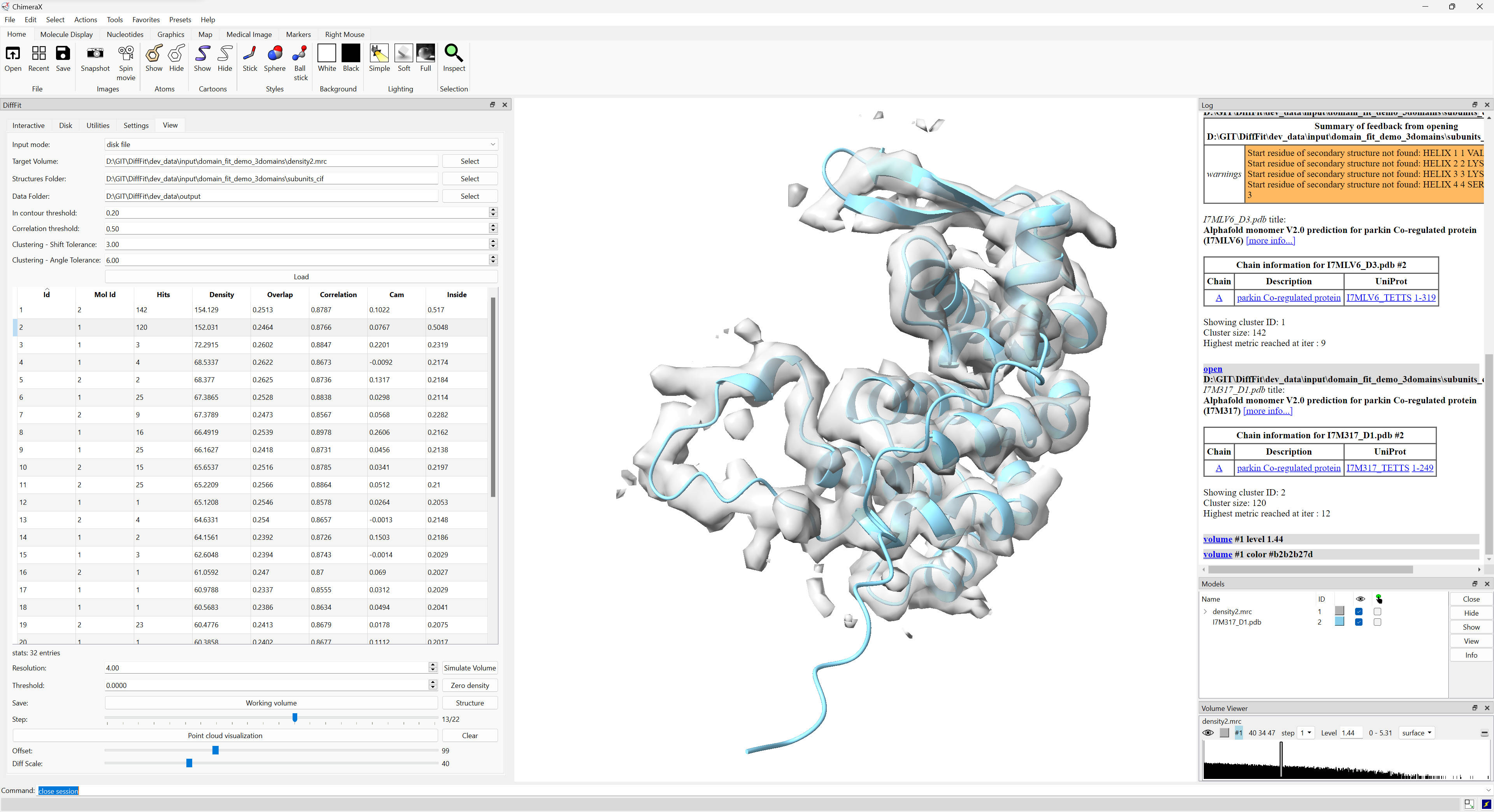}
	\caption{Visual browser based on ChimeraX. The target volume on the \change{middle} is overlaid with a fitted molecule corresponding to the selected fit result in the table on the \change{left} (clustered fits, each row is the representative placement with the highest correlation from that cluster). After inspection, users can save the placement and then select ``Simulate volume'' and ``Zero density'' to zero out the corresponding voxels from the target volume.}
	\label{fig:visual-browser}
\end{figure*}

\subsubsection{Placement of one molecule (or one subunit)}

We formulate an initial loss function $L$ to determine the best $\mathbf{p}$ and $\mathbf{q}$ parameters. This function gives us  the minimum negative average density per atom for a molecular subunit with $N$ atoms that form the set $\mathbf{X}_m$ of all atom center points T($\mathbf{x}_i) \in \mathbf{X}_m$ for a particular molecular subunit $m$:\vspace{-1ex}
\begin{equation}
L(\mathbf{p}, \mathbf{q}, \mathbf{X}_m, V) = - \left( \frac{1}{N} \sum_{i}^N D(T(\mathbf{x}_i)) \right) = - \frac{1}{N} \sum_{i}^N S(M_\mathbf{q} \cdot \mathbf{x}_{i} + \mathbf{p}, V). 
\end{equation}

We rely on the calculation of the gradient of the differentiable formulation and use it with the Adam optimizer \cite{Kingma:2017:AMS} for the optimization.
Although the Adam optimizer is known for robustness with respect to local minima, our initial loss function formulation frequently leads to a local minimum (i.e., a place that is not an optimal placement for the molecule in the map but from which the optimizer cannot find a better solution in the parameter-space neighborhood). Such local minima are a common and severe problem that also manifests in the functionality of the most commonly used tools for molecular subunit fitting (e.g., the fit-in-map feature in ChimeraX). We thus introduce several strategies to form a novel loss function , making \acronym more robust.

The first strategy that we found to substantially contribute to a good fitting performance is \textbf{filtering the input \ac{Cryo-em} map volume} $V$. For this purpose we clamp the volume values based on a user-specified minimal threshold and a minimum size of connected voxels that form a cluster. We also detect all voxels with a density value less than a given threshold and set them to zero. The size of the connected voxel cluster after thresholding must be greater than the cluster size hyperparameter. Otherwise, we set all the voxels in that cluster to zero. This step leads to the filtered volume $V_F$ and ensures that only relevant volume regions are considered for fitting, improving the focus and efficiency of the algorithm. Then, we normalize the filtered values to $[0, 1]$---a typical practice in learning and optimization approaches---, which leads to a volume $\hat{V}_F$ that turns out to be essential for controlling the magnitude of the calculations that lead to the loss function and hence the settings of the hyperparameters in the workflow.

\setlength{\lineskiplimit}{-\maxdimen}%
To accommodate the inherent noise and variability in biological datasets, we apply a series of convolution iterations to the target volume, and capture each smoothing result as a separate volume. This iterative convolutional smoothing leads to an array of volumes, and we use each of these volumes in the fitting process. This \textbf{multi-resolution approach} enhances the robustness of the fitting process by mitigating the impact of noise and data irregularities. 
Empirically, we found that a \change{3}-ele\-ment array of increasingly smoothed volumes performs well, iteratively filtered with a Gaussian smoothing kernel. We expose the size of this array as a hyperparameter to allow users to control it. We experimented with Laplacian smoothing as well, which led to unsatisfactory performance. We denote the non-smoothed volume as $\hat{V}_{F}^{G_0}$ and express the recurrent formulation of the iterative convolution smoothing as:\vspace{-1ex}
\begin{displaymath}
\hat{V}_F^{G_n} = \hat{V}_F^{G_{n-1}} * G_n.
\end{displaymath}

\setlength{\lineskiplimit}{\lineskiplimitbackup}%

\setlength{\lineskiplimit}{-\maxdimen}%
A third adaptation we apply to the initial fitting process is a stricter penalization of a mismatch. If an atom center is placed in the \ac{Cryo-em} map volume but outside the extent of the molecular target structure (i.e., outside of the target \emph{footprint}), the target density would normally be zero. To discourage such misalignment even further, we assign these regions a \textbf{negative value}. After smoothing, for voxels with a density value of zero, we replace the zero with a negative value. 
We experimented with varying the negative values or creating a smooth gradient of negative values. We noticed that a constant value of $-0.5$ outside the molecular footprint in the map performs well. We expose this value as a tunable hyperparameter. We denote the resulting volume as $\hat{V}_{F_{-c}}^{G_n}$, where $-c$ is the negative constant value. Finally, we update the loss function formulation with a volume smoothed after $j$ iterations as:\vspace{-1ex}
\begin{displaymath}
L(\mathbf{p}, \mathbf{q}, \mathbf{X}_m, \hat{V}_{F_{-c}}^{G_j}).
\end{displaymath}

\setlength{\lineskiplimit}{\lineskiplimitbackup}%

We weigh each fit with a multiresolution volume array element $w_j$ for $n$ resolutions, and sum up all the multiresolution components to form the final loss function for one $\mathbf{p}$ and $\mathbf{q}$ pair:\vspace{-1ex} 
\begin{equation}
L_m([\mathbf{p}, \mathbf{q}]) = \sum_{j=1}^{n} w_j \cdot L(\mathbf{p}, \mathbf{q}, \mathbf{X}_m, \hat{V}_{F_{-c}}^{G_j}).
\end{equation}

\setlength{\lineskiplimit}{-\maxdimen}%
To start the optimization, we need to initialize the position offset $\mathbf{p}$ and the rotation quaternion $\mathbf{q}$. We uniformly sample $N_\mathbf{q}$ points on a unit sphere and then convert these into quaternions to be applied for each offseted position. Instead of uniformly sampling positions from the volume bounding box (as in ChimeraX), we uniformly sample $N_\mathbf{p}$ positions from the positive voxels in the filtered and normalized volume $\hat{V}_{F_{-c}}$. 
This \textbf{enveloped sampling based initialization} increases the success rate by a factor of two by searching from $N_\mathbf{q} \cdot N_\mathbf{p}$ initial placements, compared to the traditional initialization in ChimeraX.

\setlength{\lineskiplimit}{\lineskiplimitbackup}%

\subsubsection{Fitting multiple placements of one molecule}

To look for fits for multiple copies of a single molecule $m$, we then take advantage of GPU parallelization and optimize all $N_\mathbf{q} \cdot N_\mathbf{p}$ pairs of $[\mathbf{p}, \mathbf{q}]$ of the molecule with atoms $\mathbf{X}_m$ altogether in one single loss function:\vspace{-1ex} 
\begin{equation}
L_{par}(m) = \sum_{k=1}^{N_\mathbf{q} \cdot N_\mathbf{p}} L_m([\mathbf{p}_k, \mathbf{q}_k]).
\end{equation}

\subsubsection{Fitting multiple placements of multiple molecules}

Finally, all subunit molecules have different numbers of atoms; it is thus not easy to parallelize the treatment of multiple molecules without overhead on the array padding of zeros. Usually, the $N_\mathbf{q} \cdot N_\mathbf{p}$ initial placements of $\mathbf{X}_m$ atoms would result in a total number of sampling operations higher than the total number of GPU threads; therefore, we process different subunits molecules sequentially in a \texttt{for} loop and form an overall loss function for $M$ molecules as follows:\vspace{-1ex}
\begin{equation}
L_{all} = \sum_{l=1}^{M} L_{par}(l).
\end{equation}

\subsubsection{Quantify the fit quality} 
\label{sec:quality-metrics}

By sampling in the simulated volume from the molecule, we can get a weight for each atom coordinate as $ W(\mathbf{x}) = S (\mathbf{x}, V_{sim})$. Then, for all  atoms in a molecule, we can form two vectors, a sampled density vector $ \mathbf{D} = [D(\mathbf{x}_1), D(\mathbf{x}_2), ..., D(\mathbf{x}_N)] $ from the target volume and a weight vector $\mathbf{W} = [W(\mathbf{x}_1), W(\mathbf{x}_2), ..., W(\mathbf{x}_N)] $ from the simulated volume. Then, we can calculate three alignment metrics, the mean overlap $\mu$, correlation $\rho$, and the correlation about the mean $\rho_\mu$ as: 
\begin{displaymath}
\mu = \frac{\mathbf{D} \cdot \mathbf{W}}{N},
\end{displaymath}

\begin{displaymath}
\rho = \frac{\mathbf{D} \cdot \mathbf{W}}{|\mathbf{D}| |\mathbf{W}|}\mathrm{, ~and}
\end{displaymath}

\begin{displaymath}
\rho_\mu = \frac{(\mathbf{D} - D_\mu) \cdot (\mathbf{W} - W_\mu)}{|\mathbf{D} - D_\mu| |\mathbf{W} - W_\mu|},
\end{displaymath}
where the subtraction operator represents subtracting the scalar average densities $D_\mu$ and $W_\mu$ from each component of the sampled density vectors. We use these quality metrics during the interactive assessment by the bioscientist in ChimeraX that we describe next.

\subsection{Visually-guided fitting}
\label{sec:Visually-guided-fitting}
A critical aspect of the post-processing of \acronym involves the clustering and sorting of the fitting results to facilitate user-guided selection and refinement. After the optimization phase, the algorithm generates a vast array of potential fits, characterized by their translation and rotation parameters. To manage this abundance of data and facilitate efficient result exploration, we apply a clustering algorithm to group the fitting results based on their spatial and orientational similarity (\autoref{fig:clustering_filtering:c}, (e)).

Each cluster represents a set of closely related fits, suggesting a consensus among them regarding the position and orientation of the fitted structure in the target volume. We sort these clusters based on a defined metric, such as the overall density overlap or correlation coefficient we just discussed, ensuring that the most promising fits are prioritized for user review. This hierarchical organization allows researchers to quickly identify the most accurate and relevant fitting results, streamlining the analysis process.

To further assist the experts in exploring the fitting results, we created an interactive visual browser as a comprehensive visualization tool to present the sorted clusters in a user-friendly format (\autoref{fig:visual-browser}). In the browser we display key metrics for each cluster, including the average correlation coefficient, the density overlap, and the consensus error measures, which provide a quick overview of the quality and relevance of each cluster. The browser also allows the biologists to select a cluster and visually inspect the fitting results within the 3D context of the target volume. This interactive exploration is crucial for assessing the fit quality in complex \ac{Cryo-em} map regions, where subtle differences in position or orientation could substantially impact the biological interpretation of the results.

\begin{figure}
    \centering
    \includegraphics[width=\columnwidth]{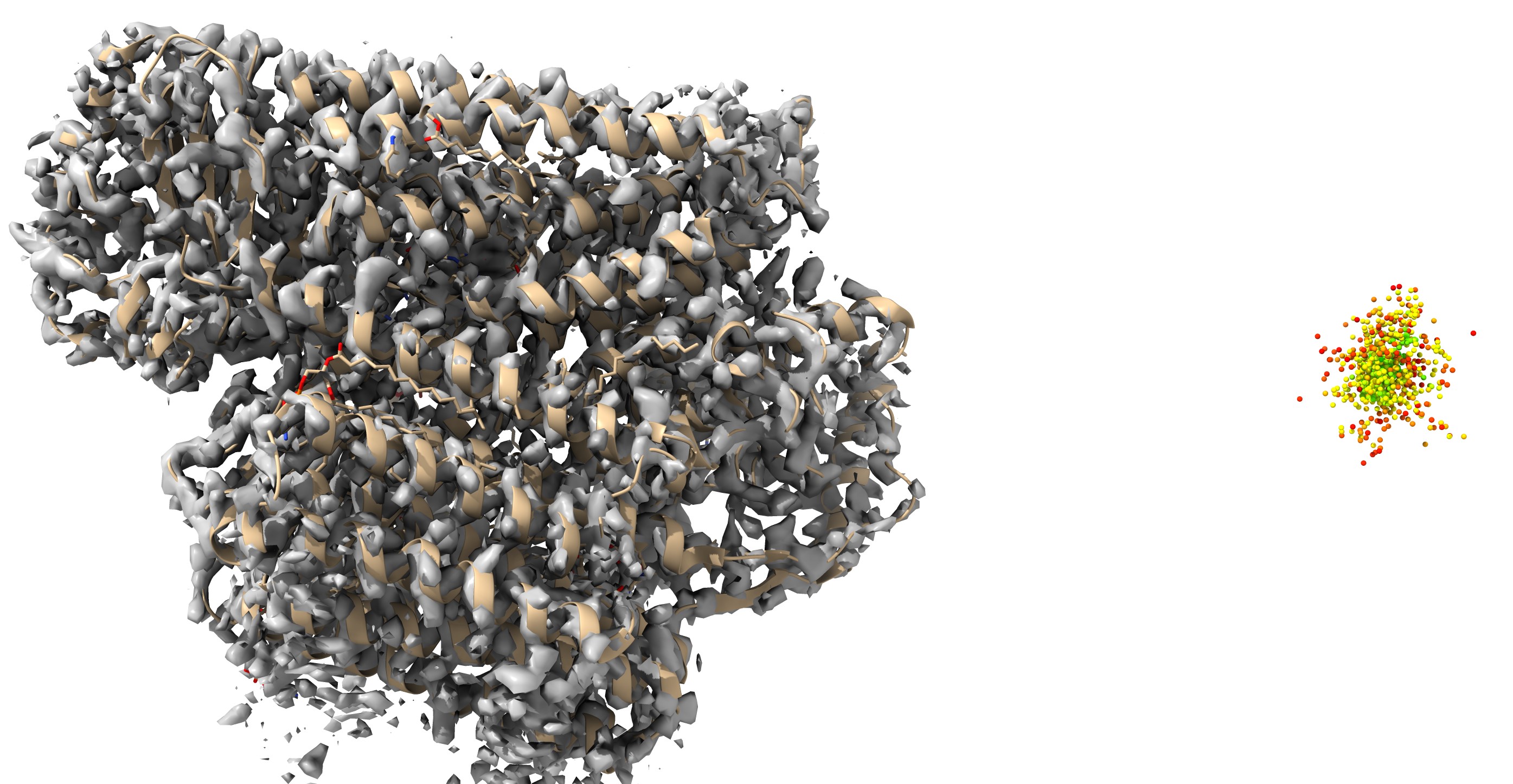}
    \caption{Interactive visual fit assessment tools we implemented in ChimeraX. Left: visualization of the best fit of the selected cluster of fits in the context of the \ac{Cryo-em} volume map; right: abstracted summary of all clusters, in which each sphere represents the center of the molecule placement and the color represents the quality of the fit. Both views are interactive and move in sync, just being offset from each other in the screen's $x$-direction (as they both represent the same spatial 3D space).}
    \label{fig:point_cloud_vis}
\end{figure}

To help the experts to gain a spatial overview of the fitting result in the tabular data (\autoref{fig:visual-browser}), we further provide means to visualize the clusters in the viewport (\autoref{fig:point_cloud_vis}). First, we display the representative molecule of each cluster of fits in the context of the \ac{Cryo-em} volume map, as we show in \autoref{fig:point_cloud_vis} on the left. The displayed fit can be controlled by selecting one from the list of fits. To also provide the experts with a visual summary of all fits, we also abstractly represent each fit cluster with the help of a sphere that we place at the center of the cluster's representative molecule as it is transformed by the fit shift and rotation (\autoref{fig:point_cloud_vis}, right). We assign each sphere a color based on the clusters' average density order and the experts can also use the spheres to select a different fit to be shown in the context of the \ac{Cryo-em} volume map.

An innovative feature of our approach is the ability to refine the fitting process iteratively by selectively excluding already placed molecules densities. Once a bioscientist selects a cluster and verifies its fit (or multiple fits) as accurate, we can zero out the corresponding density in the target volume, thus effectively removing the respective volume region from further consideration in the following part of the fitting process. 
For this purpose we, first, use the fit structure to simulate a map; then we use the voxel positions from the original map to sample the density values from the simulated map. If the sampled density is higher than a user-specified threshold, we set the original voxel's density to zero. By default we use a threshold of 0, which usually delivers good results.
This step is crucial for complex volumes containing multiple closely situated structures, as it prevents the algorithm from repeatedly fitting structures to the same volume region and reduces false positives when fitting the remaining region. 

By thus iteratively fitting and zeroing out densities, users can progressively shrink the target volume, isolating and identifying individual structures in dense or complex datasets. This iterative refinement ensures that the fitting process is not only guided by the algorithm's optimization but also by the expert's knowledge and visual assessment, ultimately leading to more accurate and biologically meaningful results. 

Our resulting visually-guided fitting framework enhances the \acronym algorithm by integrating clustering, sorting, and interactive exploration tools. These features enable users to efficiently filter through large datasets of fitting results, identify the most promising fits, and iteratively refine the fitting process based on a visual assessment. The combination of automated optimization with user-guided inspection and filtering addresses the challenge of accurately fitting molecular subunit structures within volumetric data. 

\section{Implementation}
\label{sec:implementation} 
We utilized \texttt{PyTorch} for the DiffFit algorithm implementation, capitalizing on its dynamic computational graph, automatic differentiation, and GPU acceleration to estimate positional offset and rotational quaternion parameters efficiently. Specifically, we employed \texttt{torch.nn.Conv3d} for Gaussian smoothing and \texttt{torch.nn.functional.grid\_sample} for trilinear interpolation with padding mode set to ``border.'' This approach enables us to rapidly process large volumes and multiple structures. We leveraged tensor operations and the Adam optimization algorithm for accurate optimization results. We also integrated additional functions from \texttt{SciPy}, \texttt{Bio.PDB}, \texttt{mrcfile}, and \texttt{Numpy} libraries to enhance the algorithm's functionality. In addition, we integrated \acronym seamlessly into ChimeraX (\autoref{fig:visual-browser}), based on its bundle development environment.

\section{Use case scenarios}

We designed \acronym with its advanced fitting algorithms and  integration with visualization tools to address a range of challenges in structural biology. 
To be able to illustrate their power, we now explore three key scenarios where \acronym can be particularly effective, demonstrating its versatility and potential.

\subsection{Scenario 1: Fit a single structure}
\label{sec:use-case-one}

\begin{figure}
     \centering
         \subfigure[Source structure.]{\label{fig:use_scenario_1_structure}\includegraphics[width=0.32\columnwidth]{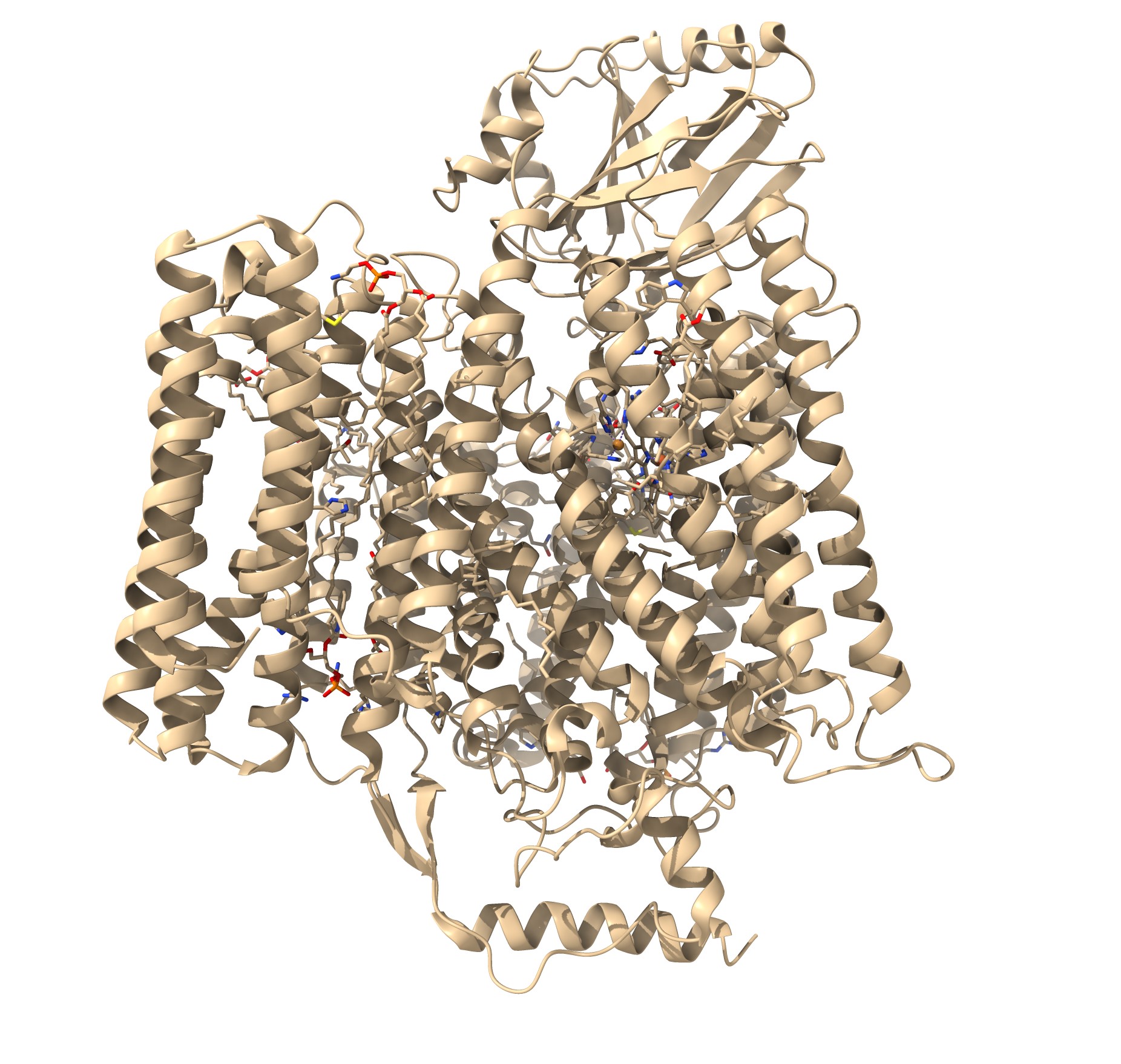}}\hfill%
         \subfigure[Target map.]{\label{fig:use_scenario_1_map}\includegraphics[width=0.32\columnwidth]{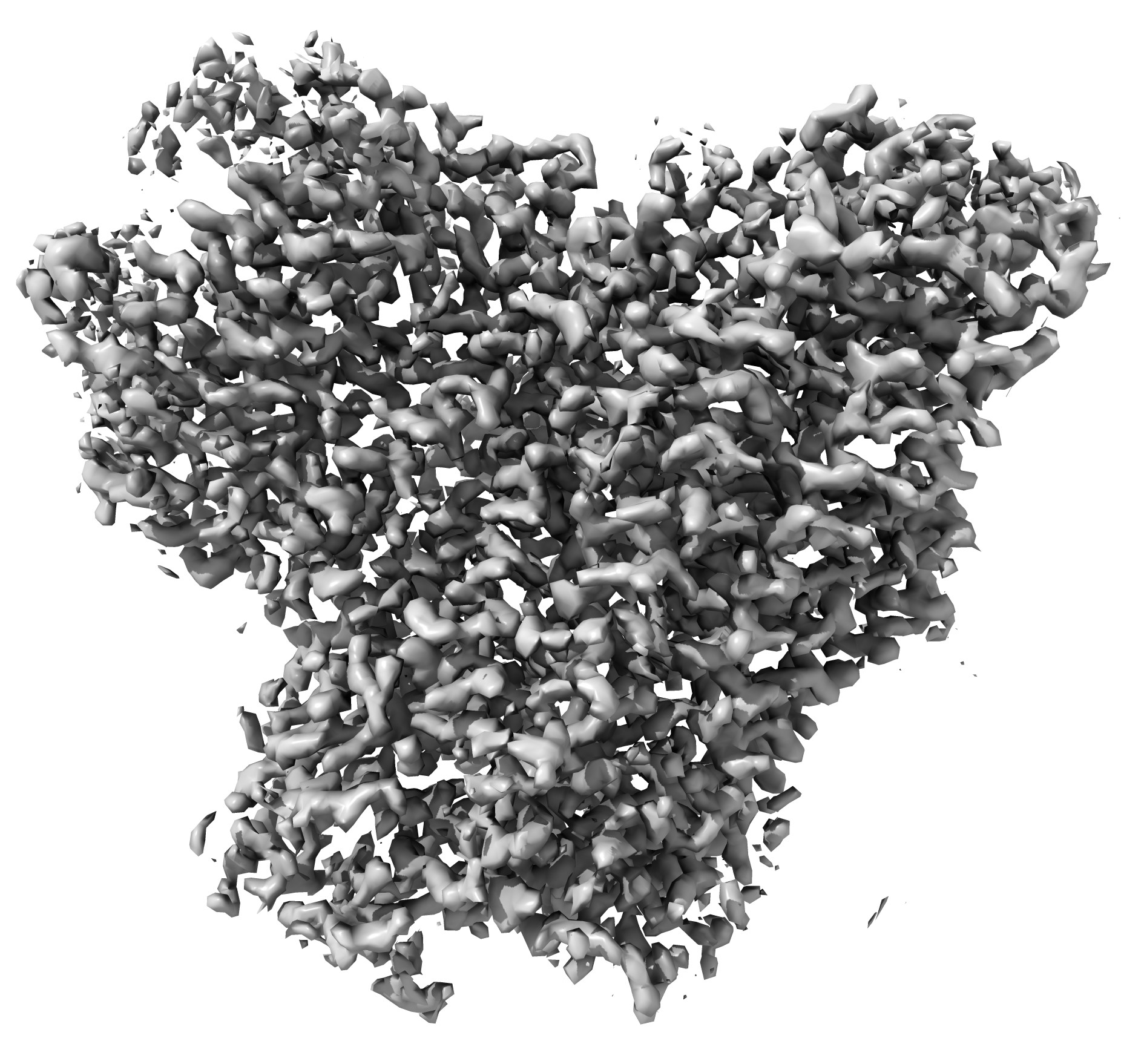}}\hfill%
         \subfigure[Fitting result.]{\label{fig:use_scenario_1_result}\includegraphics[width=0.32\columnwidth]{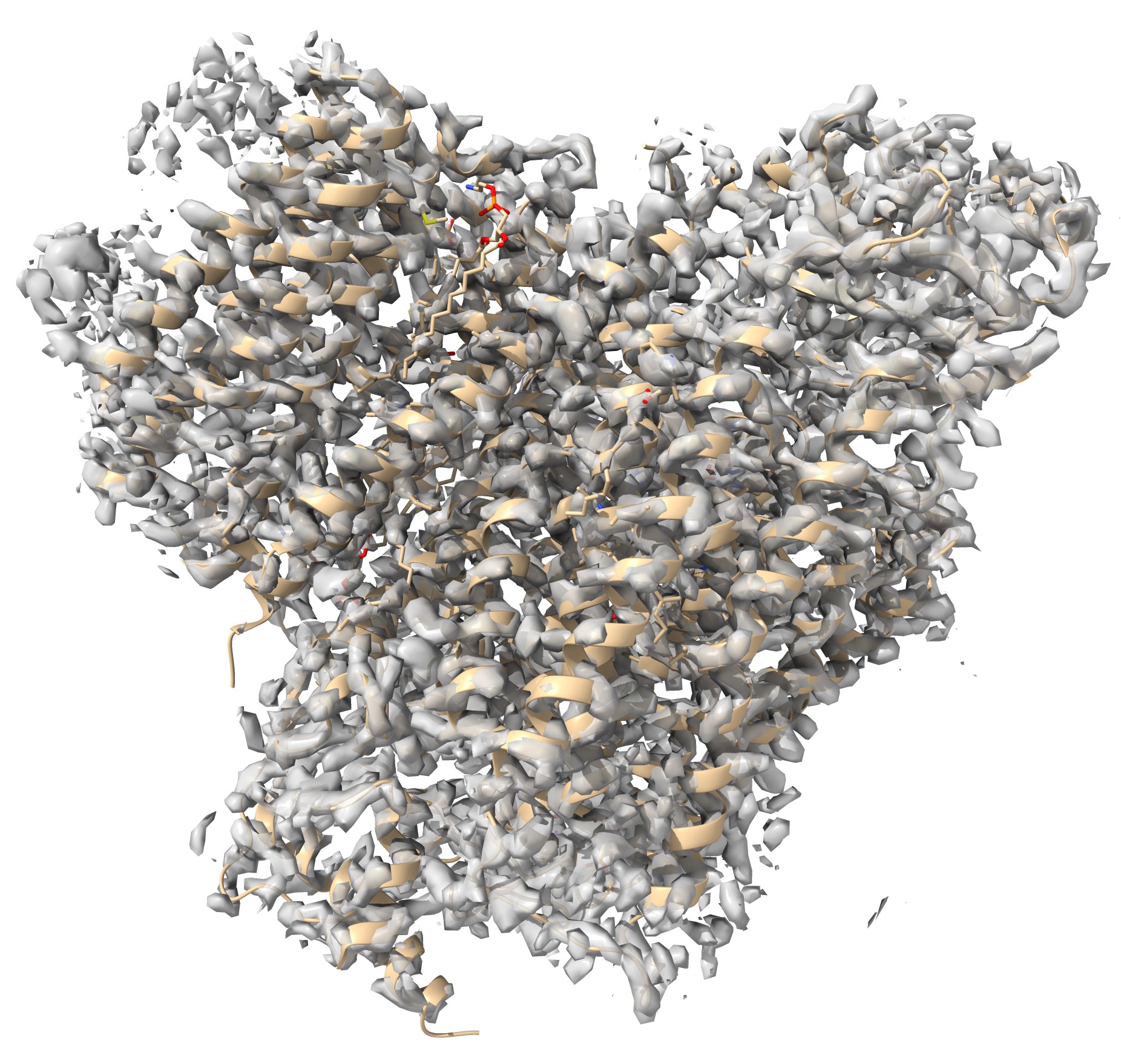}}\vspace{-2ex}
        \caption{Fitting a single structure for \href{https://doi.org/10.2210/pdb6WTI/pdb}{6WTI}.}
        \label{fig:use_scenario_1}
\end{figure}

The most straightforward application of \acronym is the fitting of a single atomistic structure to a target volumetric map, as we illustrate in \autoref{fig:use_scenario_1}. 
This scenario is common in cases in which an already-resolved protein structure or a predicted structure needs to be placed in a newly reconstructed volume captured via \ac{Cryo-em} for further refinement.

In this scenario, \acronym efficiently determines the optimal position and orientation of the protein (\autoref{fig:use_scenario_1_structure}) in the volume (\autoref{fig:use_scenario_1_map}).
The entire fitting process is automatic, removing the prerequisite of manually placing the structure at an approximate orientation close to the final optimal one. 
The interactive visually guided inspection process allows researchers to verify the fit (\autoref{fig:use_scenario_1_result}), who apply their expertise to ensure biological relevance and accuracy. 

We use the dataset reported in the recent MarkovFit work \cite{Alnabati2022} to benchmark \acronym's performance in this scenario, compare it with ChimeraX and MarkovFit, and report the results (successful hit rate, computation time, and root-mean-square deviation, RMSD) in \autoref{table:fit_single_MarkovFit_data-small} (we provide a complete table that includes each structure's EMDB ID, the number of chains, the number of atoms, voxel size, and the used surface level threshold used in \autoref{table:fit_single_MarkovFit_data} in \autoref{sec:Appendix-benchmark}). 
For each structure, we performed five experiment runs to obtain reliable results (we also attach each individual run's metrics in our supplementary material). For each run, we fit 1000\,\texttimes{} to perform the search. For ChimeraX, we ran the command ``\texttt{fit \#1 in \#2 search 1000}'' to perform atom-to-map searching, which is about one-fold faster than the atom-simulated map-to-map fitting and delivers similar results. 
We regard the number of fits in the top-ranked cluster as the hit rate if the representative fit of that cluster is within 3 Angstroms and 6 degrees (ChimeraX's default threshold) from the ground truth. We do not repeat the MarkovFit computation as it takes an average of ``7.7'' plus ``6.25'' hours (as stated in prior work \cite{Alnabati2022}) to finish a single run for each structure. Instead, we directly take the author-reported \cite{Alnabati2022} RMSD. We take the ``top-scored'' model's RMSD (although it is often the same as that of the ``best model by RMSD among top 10'') because, in practice, there is no ground truth to compare to in advance to get the best model. 
DiffFit significantly outperforms ChimeraX on the hit rate (on average, a 26.9\texttimes\ gain) and the computation time (26.0\texttimes\ gain). The RMSD of \acronym is also significantly better than MarkovFit but often slightly worse than ChimeraX's. However, this can always be corrected (the last column in the table) by a single automatic fit using ChimeraX's atom-to-map fitting, which results in a better RMSD on average. 
Compared with the overall averaged metrics, for high-resolution maps, \acronym gets higher gains on hit rate; for medium-resolution maps \acronym gets higher gains on the computation time. Of note is that, with expert knowledge, ChimeraX's performance can be boosted. For example, after smoothing the map via a command similar to ``\texttt{vol gaussian \#2 sd 2},'' the hit rate goes up to 1 (\href{https://doi.org/10.2210/pdb6WTI/pdb}{6WTI}), 67 (\href{https://doi.org/10.2210/pdb7D8X/pdb}{7D8X}), and 12 (\href{https://doi.org/10.2210/pdb6M5U/pdb}{6M5U}). 
We can also achieve a similar boosting with \acronym. As this boosting highly depends on the user's knowledge, we report only the simple version of the fitting where the user only needs to specify the surface-level threshold (which, in most cases, is ChimeraX' built-in heuristic: the top 1\% percentile of all the density values).
We computed the reported performance metrics on a workstation that uses an Nvidia RTX 4090 GPU for \acronym and a single thread on an AMD Ryzen Threadripper PRO 3995WX 2.70 GHz for ChimeraX (version 1.7.1 (2024-01-23)).


\begin{table}[t]
\centering
\setlength{\tabcolsep}{3pt}
\caption{Performance results for fitting a single structure. Res stands for resolution, C stands for ChimeraX, D stands for \acronym, M stands for MarkovFit \cite{Alnabati2022},  DC stands for DiffFit corrected by a single automatic ChimeraX fit; G stands for Gain and is D/C for Hit and C/D for Computing time (in seconds). High-avg stands for the averaged metrics for the high-resolution maps, and Med for medium maps, All for all maps.\vspace{-1ex}}
\resizebox{\columnwidth}{!}{%
\begin{tabu}{@{}lcrrrrrrrrrr@{}}
\toprule
& & \multicolumn{3}{c}{Hit rate} & \multicolumn{3}{c}{Computing time} & \multicolumn{4}{c}{RMSD (Å)} \\ 
PDB & Res & \multicolumn{3}{c}{\downbracefill} & \multicolumn{3}{c}{\downbracefill} & \multicolumn{4}{c}{\downbracefill} \\
&& \multicolumn{1}{c}{C} & \multicolumn{1}{c}{D} & \multicolumn{1}{c}{G} & \multicolumn{1}{c}{C} & \multicolumn{1}{c}{D} & \multicolumn{1}{c}{G} & \multicolumn{1}{c}{M} & \multicolumn{1}{c}{C} & \multicolumn{1}{c}{D} & \multicolumn{1}{c}{DC} \\
\midrule
\texttt{\href{https://doi.org/10.2210/pdb6WTI/pdb}{6WTI}} & 2.38 & 0.0  & \textbf{\change{136.8}} & n/a    & 150.3 & \textbf{\change{3.8}}  & \change{39.7} & 1.310 & n/a     & \change{0.942} & \textbf{\change{0.037}} \\
\texttt{\href{https://doi.org/10.2210/pdb7D8X/pdb}{7D8X}} & 2.60 & 0.0  & \textbf{\change{202.0}}  & n/a   & 196.0 & \textbf{\change{5.2}} & \change{37.6}  & 1.960 & n/a     & \change{0.984} & \textbf{0.014} \\
\texttt{\href{https://doi.org/10.2210/pdb7SP8/pdb}{7SP8}} & 2.70 & 4.6  & \textbf{\change{188}} & \change{40.9} & 130.6 & \textbf{\change{2.6}} & \change{50.5} & 1.290 & 0.996 & \change{0.969} & \textbf{\change{0.025}}  \\
\texttt{\href{https://doi.org/10.2210/pdb7STE/pdb}{7STE}} & 2.73 & 14.0 & \textbf{\change{110.4}} & \change{7.9} & 806.1 & \textbf{\change{12.1}} & \change{66.6} & 1.740 & 0.062 & \change{0.662} & \textbf{\change{0.058}} \\
\texttt{\href{https://doi.org/10.2210/pdb7JPO/pdb}{7JPO}} & 3.20 & 5.4  & \textbf{\change{191.8}} & \change{35.5} & 250.7 & \textbf{\change{6.7}} & \change{37.2}  & 2.540 & 0.017 & \change{0.922} & \textbf{0.015} \\
\texttt{\href{https://doi.org/10.2210/pdb7PM0/pdb}{7PM0}} & 3.60 & 44.0 & \textbf{\change{195.4}} & 4.4  & 352.4 & \textbf{\change{4.1}}  & \change{86.7} & 1.640 & 0.030 & \change{0.907} & \textbf{\change{0.024}}  \\
\texttt{\href{https://doi.org/10.2210/pdb6M5U/pdb}{6M5U}} & 3.80 & 0.0  & \textbf{\change{105.0}} & n/a    & 162.2 & \textbf{\change{4.1}} & \change{39.2}  & 2.360 & n/a     & \change{0.912} & \textbf{\change{0.018}} \\
\texttt{\href{https://doi.org/10.2210/pdb6MEO/pdb}{6MEO}} & 3.90 & 7.4  & \textbf{\change{116.0}} & \change{15.7} & 128.2 & \textbf{\change{3.2}}  & \change{40.1} & 1.940 & 0.489 & \change{0.786} & \textbf{0.488}  \\
\texttt{\href{https://doi.org/10.2210/pdb7MGE/pdb}{7MGE}} & 3.94 & 4.8  & \textbf{\change{123.6}} & \change{25.8} & 337.6 & \textbf{\change{4.3}} & \change{78.1} & 1.870 & \textbf{0.017} & \change{0.819} & \textbf{0.017} \\
\midrule
\multicolumn{2}{@{}l}{High-avg} & 8.9 & \textbf{\change{152.1}} & \change{21.7}  & 279.3 & \textbf{\change{5.1}} & \change{52.8} & 1.850 & 0.268 & \change{0.878} & \textbf{0.077} \\
\midrule
\texttt{\href{https://doi.org/10.2210/pdb5NL2/pdb}{5NL2}} & 6.60  & 1.8   & \textbf{\change{163.2}} & \change{90.7} & 94.6   & \textbf{\change{2.0}}  & \change{48.0}  & 2.440  & 0.093 & \change{1.124} & \textbf{\change{0.056}} \\
\texttt{\href{https://doi.org/10.2210/pdb7K2V/pdb}{7K2V}} & 6.60  & 49.0  & \textbf{\change{165.6}} & \change{3.4}  & 240.6  & \textbf{\change{4.1}}  & \change{58.2} & 25.290 & \textbf{0.338} & \change{1.323} & \textbf{\change{0.338}} \\
\texttt{\href{https://doi.org/10.2210/pdb7CA5/pdb}{7CA5}} & 7.60  & 55.8  & \textbf{\change{72.4}} & \change{1.3}  & 322.6  & \textbf{\change{2.9}}  & \change{110.0} & 3.290  & 2.042 & \textbf{1.207} & \change{2.042} \\
\texttt{\href{https://doi.org/10.2210/pdb5VH9/pdb}{5VH9}} & 7.70  & 68.6  & \textbf{\change{158.0}} & \change{2.3}  & 1147.8 & \textbf{\change{14.1}} & \change{81.3} & 0.960  & \textbf{0.085} & \change{0.991} & \textbf{\change{0.085}} \\
\texttt{\href{https://doi.org/10.2210/pdb6AR6/pdb}{6AR6}} & 9.00  & 78.0  & \textbf{\change{182.6}} & \change{2.3}  & 74.9   & \textbf{\change{1.5}}  & \change{49.3} & 2.200  & 0.123 & \change{2.617} & \textbf{\change{0.117}} \\
\texttt{\href{https://doi.org/10.2210/pdb3J1Z/pdb}{3J1Z}} & 13.00  & 138.6 & \textbf{\change{172.2}} & \change{1.2}  & 64.4   & \textbf{\change{2.0}}  & \change{33.0} & 32.330 & 0.396 & \change{2.612} & \textbf{\change{0.388}} \\
\midrule
\multicolumn{2}{@{}l}{Med-avg} & 65.3   & \textbf{\change{152.3}} & \change{16.9}  & 324.1 & \textbf{\change{4.4}}    & \change{63.3} & 11.085 & 0.513  & \change{1.646} & \textbf{\change{0.504}} \\
\midrule
\multicolumn{2}{@{}l}{All-avg} & 31.5   & \textbf{\change{152.2}} & \change{19.8}  & 297.3 & \textbf{\change{4.9}}   & \change{57.0} & 5.544  & 0.366  & \change{1.185} & \textbf{\change{0.248}} \\
\bottomrule
\end{tabu}}
\label{table:fit_single_MarkovFit_data-small}
\end{table}

\subsection{Scenario 2: Composite multiple structures}

A more complex use case involves the fitting of multiple structures to a single, large, often complex, volumetric dataset, such as assembling a viral capsid from individual protein units or constructing a ribosomal complex from its constituent proteins and RNA molecules. \acronym can handle such composite fitting tasks by iteratively optimizing the placement of each component, from largest to smallest, while considering the spatial relationships and interactions between them to prevent overlaps and ensure a coherent assembly.

In \autoref{fig:teaser} we show an example of compositing multiple structures into a \ac{Cryo-em} volume map, for the PDB-protein \href{https://doi.org/10.2210/pdb8SMK/pdb}{8SMK} \cite{Zhou:2024:ADI}. In the first row we demonstrate how the middle and bottom parts are fitted in the first computation round, with the remaining top part of the protein being fitted in the second round of the interactive process.

The ability to \emph{zero-out} densities, once a fit is interactively confirmed by the expert, allows \acronym to fit multiple structures sequentially without interference from previously placed components, as we demonstrate in \autoref{fig:teaser}.  This iterative approach is particularly useful for densely packed molecular complexes, where individual components may be difficult to distinguish in the volumetric data. This scenario is critical for understanding the functional context of proteins in larger biomolecular assemblies or cellular environments.

\subsection{Scenario 3: Identify unknown densities}

\acronym also offers bioscientists the capability to identify and characterize unknown densities within volumetric datasets. In cases where a volume contains unassigned or ambiguous regions, possibly indicating the presence of previously unidentified molecules or molecular complexes, \acronym can be used to screen a library of known structures and predicted structures for potential matches.

By fitting the structures from a library to the unidentified densities and evaluating the fit quality, researchers can hypothesize the identity of the unknown components. This scenario is invaluable for discovery-based research, where identifying novel components in complex molecular assemblies could lead to significant biological insights.

\begin{figure}
     \centering
         \subfigure[Library of structures to search against (subset).]{\label{fig:use_scenario_3_library}\includegraphics[width=\columnwidth]{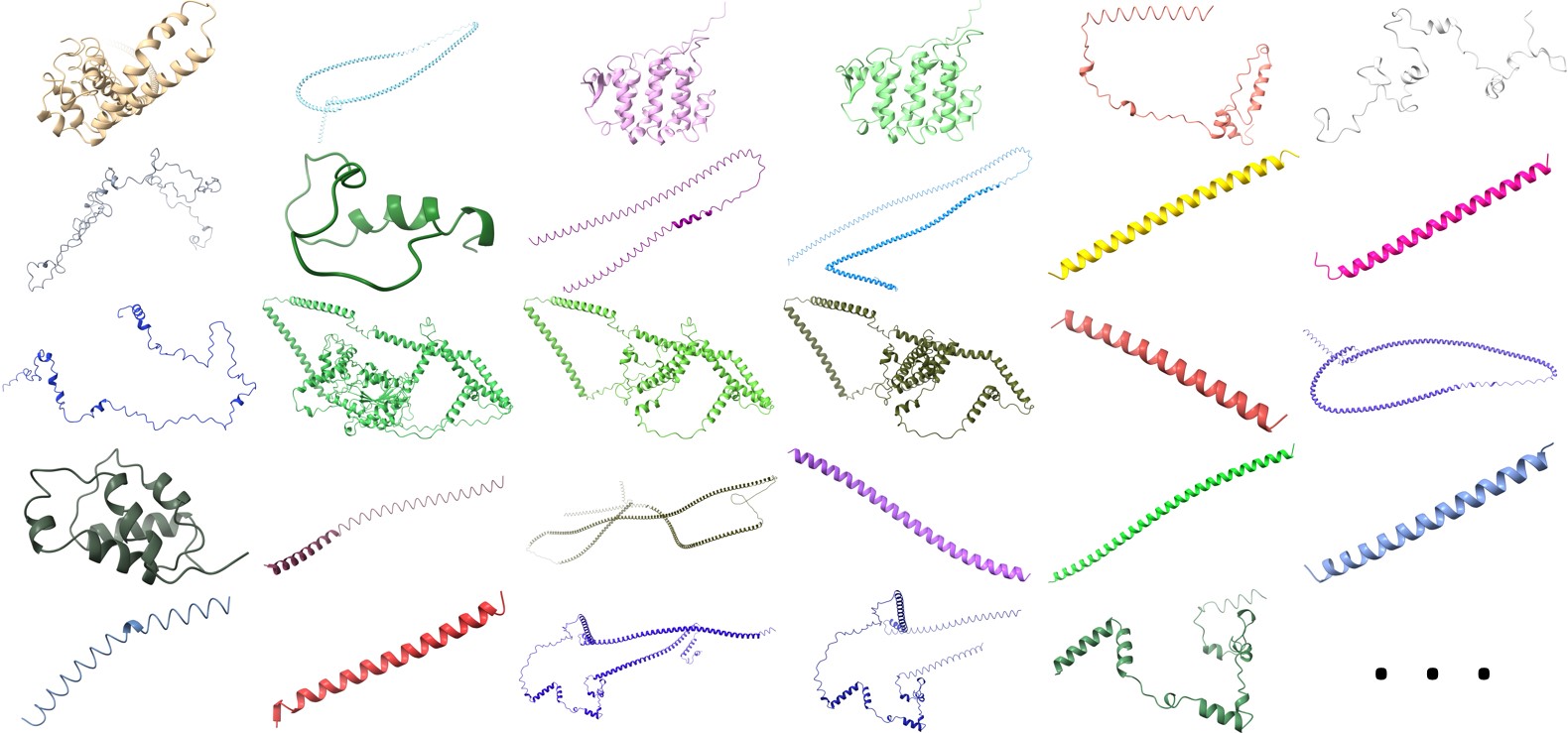}}\\[1ex]
         \subfigure[Unknown density identification: comparison of three potential fits which are overlaid on top of the target volumes.]{\label{fig:use_scenario_3_result}\includegraphics[width=\columnwidth]{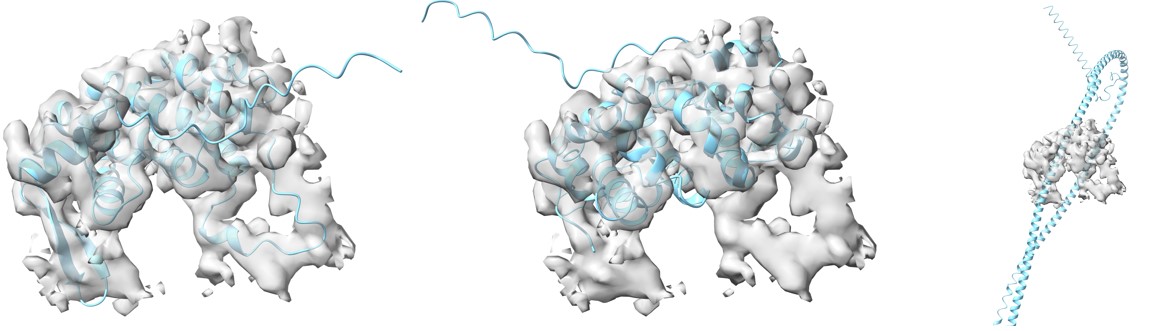}}\vspace{-2ex}
        \caption{Unknown density identification, where dozens to hundreds of molecular structures can be evaluated for potential fit.}\vspace{2ex}
        \label{fig:three graphs}
\end{figure}

We performed a search with the demonstration dataset from a recent automated domain-level protein identification technique, DomainFit \cite{Gao:2023:DIP}. The task was to identify, from a library of 359 protein domains (\autoref{fig:use_scenario_3_library} shows a subset), which domains best fits into the given volume (\autoref{fig:use_scenario_3_result}). DomainFit identified two candidates (I7MLV6\_D3, and I7M317\_D1), after first fitting all domains using the \texttt{fitmap} command in ChimeraX and then performing statistical analyses to remove false positives. With our \acronym, we identified these same two candidates with our fitting technique followed by a visual inspection, where the two candidates were the first to be inspected. We report the hit rates in \autoref{table:identify_unknown}, which shows a $\approx$\,2\texttimes\ gain on the hit rate. In addition, DomainFit takes $\approx$12 hours to finish the task, while with \acronym it takes $\approx$\,\change{7} minutes---a \change{103}\texttimes\ gain in computation time.

\begin{table}[t]
\caption{Performance results for identifying unknown structures. C stands for ChimeraX, D stands for \acronym; Gain = D/C for Hit.\vspace{-1ex}}
\centering
\begin{tabu}{llll}
\toprule
Structure  & C Hit & D Hit & Gain \\
\midrule
I7MLV6\_D3 & 108   & \change{254}   & \change{2.4}\texttimes  \\
I7M317\_D1 & 127   & \change{240}   & \change{1.9}\texttimes  \\
\bottomrule
\end{tabu}
\label{table:identify_unknown}
\end{table}

\section{Feedback}

In addition to these quantitative metrics, qualitative feedback from users plays a crucial role in evaluating the practical utility and user experience of \acronym. We thus solicited feedback from a diverse group of users, including PhD students, structural biologists, and computational scientists; through surveys, interviews, and hands-on testing sessions.

Specifically, we emailed \ac{Cryo-em} practitioners at our university to test our tool. We performed three Zoom-based demo sessions and one in-person demo and feedback session with those people who expressed interest. The first session involved five people from a research group that studies structural biology and engineering (one senior PhD student specializing in structural and computational biology, one senior PhD student in experimental biology about protein structures, one senior PhD student in experiments elucidating the structure and function of proteins, one postdoc in experimental biology about protein structures involved in cell signaling, one research scientist in using Cryo-EM and other techniques to study protein structures). The second session involved two people from a research group focusing on using \ac{Cryo-em} to study the structural biology of human DNA replication and repair (an assistant professor and his postdoc). The third session involved one junior PhD student using \ac{Cryo-em} to solve protein structures. The final in-person session involved one research scientist who provides \ac{Cryo-em} service as a platform to the whole university and builds a bridge between the microscope and users at all levels.
Apart from verbal feedback during and after these demo sessions, one participant in the first group, as well as the participant in the in-person session, also sent us written feedback, which we include in \autoref{sec:Appendix-feedback}.
In total, nine people provided qualitative feedback, which we report next. 

\textbf{Usability.} The participants appreciated \acronym's intuitive interface and workflow (``quite intuitive and easy to use'') and the direct integration into ChimeraX with its known UI, which significantly lowers the barrier to entry, in particular, for new users, while still providing advanced features for experienced researchers. In particular the easy one-step fitting of a PDB file into a \ac{Cryo-em} map was highlighted. The visually guided fitting process was appreciated as a powerful feature for refining fitting results based on expert judgment.  Suggestions for further improvements of this interaction step and the UI in general were to display several visual clusters at a time, to visually compare the fit results, and to offer a tab-based method to check the various clusters. This approach would make it easier to observe how each cluster fits relative to the \ac{Cryo-em} volume; thus, researchers would be better able to focus on a chain of interest. We have addressed this point already with our visual fit assessment tools (\autoref{fig:point_cloud_vis} and \autoref{sec:Visually-guided-fitting}). One issue we noticed is that some experts commented on the computation times in the order of a few minutes, which points to the fact that they did not have a CUDA-compatible GPU at their disposal. As we demonstrated with our runtime analysis in \autoref{sec:use-case-one}, with such affordable hardware, the computation is possible in well under a minute for many structures, so \ac{Cryo-em} labs can break free from extremely long computation times or the need for employing large (and expensive) computation resources.

\textbf{Impact on Research.} The participants reported that \acronym has the potential to have a tangible impact on their research, enabling them to address complex fitting challenges that were previously out of reach. They thus get the ability to rapidly sample a large database of candidate structures for regions with unassigned protein density. The new ability to fit structures rapidly and accurately  into volumetric data offers new directions of investigation and has the potential to accelerate discoveries in structural biology. One respondent stated that our automatic fitting and visual inspection approach ``could be a key feature in ChimeraX that [could become] a standard in many pipelines'' as well as ``a key implementation [for] a standard modeling workflow.''  

The respondents also made suggestions for future developments such as better structuring the handling of the associated files, further exploring the design space of the abstract cluster representations, a detailed protocol for downloading and installing the tool (already implemented in our GitHub repository), as well as adding workflows that would require automatically creating model subdivisions and performing fits on those before integrating the results into the reference volume. They also highlighted further potential uses, such as employing our approach to analyze density maps generated using X-ray crystallography.

\section{Limitations and Future Work}
Although our \acronym technique demonstrates a substantial improvement over state-of-the-art techniques, it also has some limitations.
One issue arises in Scenario 2, due to the gradual \emph{zeroing-out} of the target map, as multiple structures are being composited onto a single map.
In this case, the volume removal can potentially remove some voxels that are also part of adjacent interacting molecular subunits, especially in those cases when the chosen fit is not exactly perfect. Then, the subsequent subunit to be placed may have a higher difficulty of finding its correct location because these few zeroed voxels penalize the correct position. We are currently investigating this limitation and plan to experiment with heuristics that remove only voxels that are fully covered by the molecular structure or slightly shrink (by using a higher user-specified threshold) the to-be-\emph{zeroed-out} region before removing it from the map. Another related limitation is that the fitting process does not check for collisions in adjacent subunits, as previously investigated in visualization work (\eg, \cite{Schmidt:2014:YMCA}). While the map shrinking approach should address this issue, subunits may still overlap in some cases. Resolving this issue is straightforward: we could remove those subunit poses that overlap with previously confirmed subunit placements or could minimally modify the pose to resolve the structural collisions.

Our technique is also not entirely parameter-free. The threshold value for removal of voxels with small density values, \eg, and the threshold value for the connected-voxel minimal cluster size have to be manually set. Manipulating these parameters requires prior domain experience with \ac{Cryo-em} data from the bioscientists. We plan to automate the identification of these thresholds or, at least, define suitable default values and provide guidance on reasonable parameter ranges.

The visualization design in our current version of the tool also offers only the most essential visual encoding types, with a clear potential for further improvement. We plan to adapt comparative visualization techniques for intersecting surfaces and smart visibility techniques for combined rendering of the volumetric density representation with molecular surfaces or cartoon representations. Furthermore, the visualization design for analyzing the structural poses in a cluster suffers from high spatial occlusion among the subunits' various poses (see \autoref{fig:clustering_filtering:d}).  This is a common domain problem and requires a dedicated research effort to combat the occlusion of this magnitude. Our tool can also be further extended in future work to, for instance, change the color encoding of the fitting results presentation to use one of the other alignment metrics (\autoref{sec:quality-metrics}) or to change the size of the spheres in the results summary to make larger clusters appear more prominent.

Driven by the feedback we received, additional possible future research directions include extending \acronym to deformable transformations,  considering stoichiometry and symmetry, and speeding it up even further through an integration with CUDA. 

\section{Conclusion}
\label{sec:conclusion}
In conclusion, \acronym offers a novel and efficient solution to the challenges of atom-to-map fitting that has arisen in structural biology. We address these challenges with our differentiable fitting along with a set of essential strategies that make our algorithm robust for the domain problem. Our approach adds a human-in-the-loop visual analytics approach to the workflow and provides an open-source package designed to work as part of a standard software tool on the domain (ChimeraX).
A key element of our approach is the change from pixel-to-pixel (or the equivalent, voxel-to-voxel) fitting to point-to-volume fitting, which enabled us to deal with the specific constraints of the fitting problem in structural biology. 
In hindsight, as we witness the success of using atom coordinates to sample the volume instead of using all the voxels, DiffFit suggests that, in Reddy's \cite{Reddy2020Differentiable} original technique that inspired our work, maybe not all the pixels in a patch are needed for the optimization, which means that
the image compositing optimization could be more efficient and could
handle larger examples with faster processing.


The quantitative performance metrics we reported collectively demonstrate \acronym's substantial improvement over traditional approaches, highlighting its potential to transform the field of structural biology. This potential is particularly large because the workflows of Assembline, a protocol published in 2022 \cite{Rantos:2022:ISM}, and DomainFit \cite{Gao:2023:DIP}, so far, rely on ChimeraX's fitmap command as their first step. This stage can now be replaced by our more effective \acronym, laying a new foundation for these and other workflows. By demonstrating the effectiveness of our technique across three scenarios, from fitting individual structures to assembling complex molecular architectures and identifying unknown components, we demonstrated that \acronym overcomes the limitations of manually placing individual molecules before fitting, makes the compositing faster, more accurate, and intuitive, and opens the possibility of scanning the whole set of known and predicted molecules with the current computational resources. Ultimately, we thus escape the \href{https://en.wiktionary.org/wiki/Faustian_bargain}{Faustian bargain} \cite{Goethe:1808:F} and instead are now free ourselves to explore the inner workings of the biological world.


\acknowledgments{%
This research was supported by King Abdullah University of Science and Technology (KAUST) (BAS/1/1680-01-01). 
We thank Tom Goddard for his ChimeraX development support, all evaluators for providing their feedback, \change{Alexandra Irger for her narration in the submission video,} Nagarajan Kathiresan for his supercomputer support, and Rahman Hasan for his hardware support. D. Luo thanks Todd Pietruszka, Irfan Anjum, Rand Qoj, Xinwei Qiu, Yonghong Leng, Yujing Ouyang, Ziyun Zhang, and Xiaode Zhu for their kind support.%
}

\section*{Supplemental material pointers}

We provide the open-source repository of our tool at \href{https://github.com/nanovis/DiffFit}{\texttt{github\discretionary{}{.}{.}com\discretionary{/}{}{/}nanovis\discretionary{/}{}{/}DiffFit}}. Other supplemental material is available at \href{https://osf.io/5tx4q/}{\texttt{osf.io/5tx4q}}.

\section*{Images license and copyright}
We as authors state that all of our figures are and remain under our own personal copyright, with the permission to be used here. We also make them available under the \href{https://creativecommons.org/licenses/by/4.0/}{Creative Commons At\-tri\-bu\-tion 4.0 International (\ccLogo\,\ccAttribution\ \mbox{CC BY 4.0})} license and share them at \href{https://osf.io/5tx4q/}{\texttt{osf.io/5tx4q}}.

\bibliographystyle{abbrv-doi-hyperref}
\bibliography{abbreviations,references}

\clearpage

\appendix 

\begin{strip} 
\noindent\begin{minipage}{\textwidth}
\makeatletter
\centering%
\sffamily\bfseries\fontsize{15}{16.5}\selectfont
\mytitle\\[.5em]
\large Appendix\\[.75em]
\makeatother
\normalfont\rmfamily\normalsize\noindent\raggedright In this appendix we provide additional explanations, tables, plots, and charts that show data beyond the material that we could include in the main paper due to space limitations or because it was not essential for explaining our approach.
\end{minipage}
\end{strip}
\section{Detailed benchmark results for use case scenario 1---Fit a single structure}
\label{sec:Appendix-benchmark}

In \autoref{table:fit_single_MarkovFit_data} we provide a detailed benchmark table for the first use case.

\begin{table}[b!]
\begin{minipage}{\textwidth}
\centering
\setlength{\tabcolsep}{2.5pt}
\caption{\label{table:fit_single_MarkovFit_data}Performance results for fitting a single structure. S stands for subunits, A stands for atoms, Res stands for resolution (in Å), Vs stands for voxel size, L stands for surface level threshold, C stands for ChimeraX, D stands for \acronym, M stands for MarkovFit \cite{Alnabati2022},  DC stands for DiffFit corrected by a single automatic ChimeraX fit; G stands for Gain and is D/C for Hit and C/D for Computing time (in seconds).\vspace{-1ex}}
\begin{tabu}{@{}lrrrrrr@{\hspace{15pt}}rrr@{\hspace{15pt}}rrr@{\hspace{15pt}}rrrr@{}}
\toprule

&&&&&&& \multicolumn{3}{c}{Hit rate} & \multicolumn{3}{c}{Computing time} & \multicolumn{4}{c}{RMSD (Å)} \\ 
PDB & EMDB & \#S & \#A & Res & Vs & L & \multicolumn{3}{c}{\downbracefill} & \multicolumn{3}{c}{\downbracefill} & \multicolumn{4}{c}{\downbracefill} \\
&&&&&&& \multicolumn{1}{c}{C} & \multicolumn{1}{c}{D} & \multicolumn{1}{c}{G} & \multicolumn{1}{c}{C} & \multicolumn{1}{c}{D} & \multicolumn{1}{c}{G} & \multicolumn{1}{c}{M} & \multicolumn{1}{c}{C} & \multicolumn{1}{c}{D} & \multicolumn{1}{c}{DC} \\

\midrule
\texttt{\href{https://doi.org/10.2210/pdb6WTI/pdb}{6WTI}} & 21897 & 4 & 9,980  & 2.38 & 1.08 & 0.7660 & 0.0  & \textbf{\change{136.8}} & n/a    & 150.3 & \textbf{\change{3.8}}  & \change{39.7} & 1.310 & n/a     & \change{0.942} & \textbf{\change{0.037}} \\
\texttt{\href{https://doi.org/10.2210/pdb7D8X/pdb}{7D8X}} & 30614 & 4 & 10,928 & 2.60 & 1.08 & 0.0229 & 0.0  & \textbf{\change{202.0}}  & n/a   & 196.0 & \textbf{\change{5.2}} & \change{37.6}  & 1.960 & n/a     & \change{0.984} & \textbf{0.014} \\
\texttt{\href{https://doi.org/10.2210/pdb7SP8/pdb}{7SP8}} & 25368 & 3 & 6,090 & 2.70 & 1.08 & 5.5755 & 4.6  & \textbf{\change{188}} & \change{40.9} & 130.6 & \textbf{\change{2.6}} & \change{50.5} & 1.290 & 0.996 & \change{0.969} & \textbf{\change{0.025}}  \\
\texttt{\href{https://doi.org/10.2210/pdb7STE/pdb}{7STE}} & 25426 & 5 & 14,249 & 2.73 & 0.83 & 0.0963 & 14.0 & \textbf{\change{110.4}} & \change{7.9} & 806.1 & \textbf{\change{12.1}} & \change{66.6} & 1.740 & 0.062 & \change{0.662} & \textbf{\change{0.058}} \\
\texttt{\href{https://doi.org/10.2210/pdb7JPO/pdb}{7JPO}} & 22417 & 5 & 16,087 & 3.20 & 1.07 & 0.0240 & 5.4  & \textbf{\change{191.8}} & \change{35.5} & 250.7 & \textbf{\change{6.7}} & \change{37.2}  & 2.540 & 0.017 & \change{0.922} & \textbf{0.015} \\
\texttt{\href{https://doi.org/10.2210/pdb7PM0/pdb}{7PM0}} & 13508 & 3 & 10,169 & 3.60 & 1.10 & \change{0.0068} & 44.0 & \textbf{\change{195.4}} & 4.4  & 352.4 & \textbf{\change{4.1}}  & \change{86.7} & 1.640 & 0.030 & \change{0.907} & \textbf{\change{0.024}}  \\
\texttt{\href{https://doi.org/10.2210/pdb6M5U/pdb}{6M5U}} & 30093 & 3 & 10,549 & 3.80 & 1.06 & 0.0350 & 0.0  & \textbf{\change{105.0}} & n/a    & 162.2 & \textbf{\change{4.1}} & \change{39.2}  & 2.360 & n/a     & \change{0.912} & \textbf{\change{0.018}} \\
\texttt{\href{https://doi.org/10.2210/pdb6MEO/pdb}{6MEO}} & 9108  & 3 & 7,465  & 3.90 & 1.06 & 0.0500 & 7.4  & \textbf{\change{116.0}} & \change{15.7} & 128.2 & \textbf{\change{3.2}}  & \change{40.1} & 1.940 & 0.489 & \change{0.786} & \textbf{0.488}  \\
\texttt{\href{https://doi.org/10.2210/pdb7MGE/pdb}{7MGE}} & 23827 & 4 & 9,010  & 3.94 & 0.94 & 0.2550 & 4.8  & \textbf{\change{123.6}} & \change{25.8} & 337.6 & \textbf{\change{4.3}} & \change{78.1} & 1.870 & \textbf{0.017} & \change{0.819} & \textbf{0.017} \\

\midrule
\multicolumn{2}{@{}l}{\change{High-avg}} & 3.78 & 10,503  & 3.21 & 1.03 & n/a & 8.9 & \textbf{\change{152.1}} & \change{21.7}  & 279.3 & \textbf{\change{5.1}} & \change{52.8} & 1.850 & 0.268 & \change{0.878} & \textbf{0.077} \\
\midrule

\texttt{\href{https://doi.org/10.2210/pdb5NL2/pdb}{5NL2}} & 3658   & 2 & 4,312  & 6.60 & 1.35 & 0.0297 & 1.8   & \textbf{\change{163.2}} & \change{90.7} & 94.6   & \textbf{\change{2.0}}  & \change{48.0}  & 2.440  & 0.093 & \change{1.124} & \textbf{\change{0.056}} \\
\texttt{\href{https://doi.org/10.2210/pdb7K2V/pdb}{7K2V}} & 22647  & 2 & 5,717  & 6.60 & 1.05 & 0.0050 & 49.0  & \textbf{\change{165.6}} & \change{3.4}  & 240.6  & \textbf{\change{4.1}}  & \change{58.2} & 25.290 & \textbf{0.338} & \change{1.323} & \textbf{\change{0.338}} \\
\texttt{\href{https://doi.org/10.2210/pdb7CA5/pdb}{7CA5}} & 30324  & 2 & 6,484  & 7.60 & 1.06 & 0.0100 & 55.8  & \textbf{\change{72.4}} & \change{1.3}  & 322.6  & \textbf{\change{2.9}}  & \change{110.0} & 3.290  & 2.042 & \textbf{1.207} & \change{2.042} \\
\texttt{\href{https://doi.org/10.2210/pdb5VH9/pdb}{5VH9}} & 8673   & 2 & 22,042 & 7.70 & 1.20 & 0.0074 & 68.6  & \textbf{\change{158.0}} & \change{2.3}  & 1147.8 & \textbf{\change{14.1}} & \change{81.3} & 0.960  & \textbf{0.085} & \change{0.991} & \textbf{\change{0.085}} \\
\texttt{\href{https://doi.org/10.2210/pdb6AR6/pdb}{6AR6}} & 8898   & 2 & 2,395  & 9.00  & 3.00 & 0.0739 & 78.0  & \textbf{\change{182.6}} & \change{2.3}  & 74.9   & \textbf{\change{1.5}}  & \change{49.3} & 2.200  & 0.123 & \change{2.617} & \textbf{\change{0.117}} \\
\texttt{\href{https://doi.org/10.2210/pdb3J1Z/pdb}{3J1Z}} & 5450   & 2 & 4,586  & 13.00 & 2.74 & 3.8989 & 138.6 & \textbf{\change{172.2}} & \change{1.2}  & 64.4   & \textbf{\change{2.0}}  & \change{33.0} & 32.330 & 0.396 & \change{2.612} & \textbf{\change{0.388}} \\


\midrule
\multicolumn{2}{@{}l}{\change{Med-avg}} & 2.00 & 7,589  & 8.42 & 1.73 & n/a & 65.3   & \textbf{\change{152.3}} & \change{16.9}  & 324.1 & \textbf{\change{4.4}}    & \change{63.3} & 11.085 & 0.513  & \change{1.646} & \textbf{\change{0.504}} \\
\midrule
\multicolumn{2}{@{}l}{\change{All-avg}} & 3.01 & 9,337  & 5.29 & 1.31 & n/a & 31.5   & \textbf{\change{152.2}} & \change{19.8}  & 297.3 & \textbf{\change{4.9}}   & \change{57.0} & 5.544  & 0.366  & \change{1.185} & \textbf{\change{0.248}} \\

\bottomrule
\end{tabu}
\end{minipage}
\end{table}

\section{Details on the user feedback sessions}
\label{sec:Appendix-feedback}

One participant in the first feedback group sent us written feedback in addition to the comments during the Zoom session, which we attach in anonymized form at the very end of this appendix. In addition, the expert who participated in the in-person session also sent additional feedback by e-mail, which we also include in anonymized form at the end of the appendix.

\clearpage

\includepdf[pages=-]{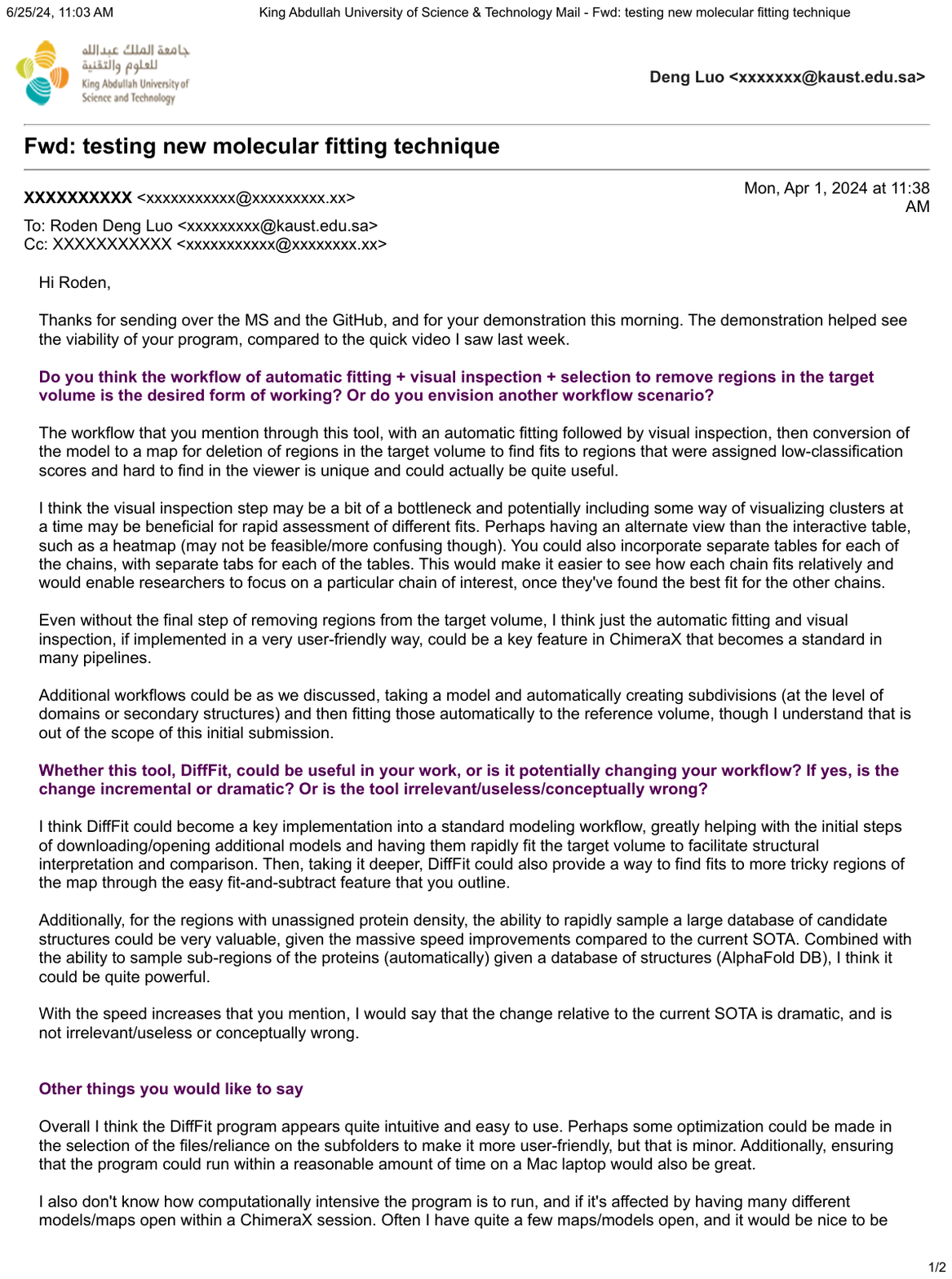}
\includepdf[pages=-]{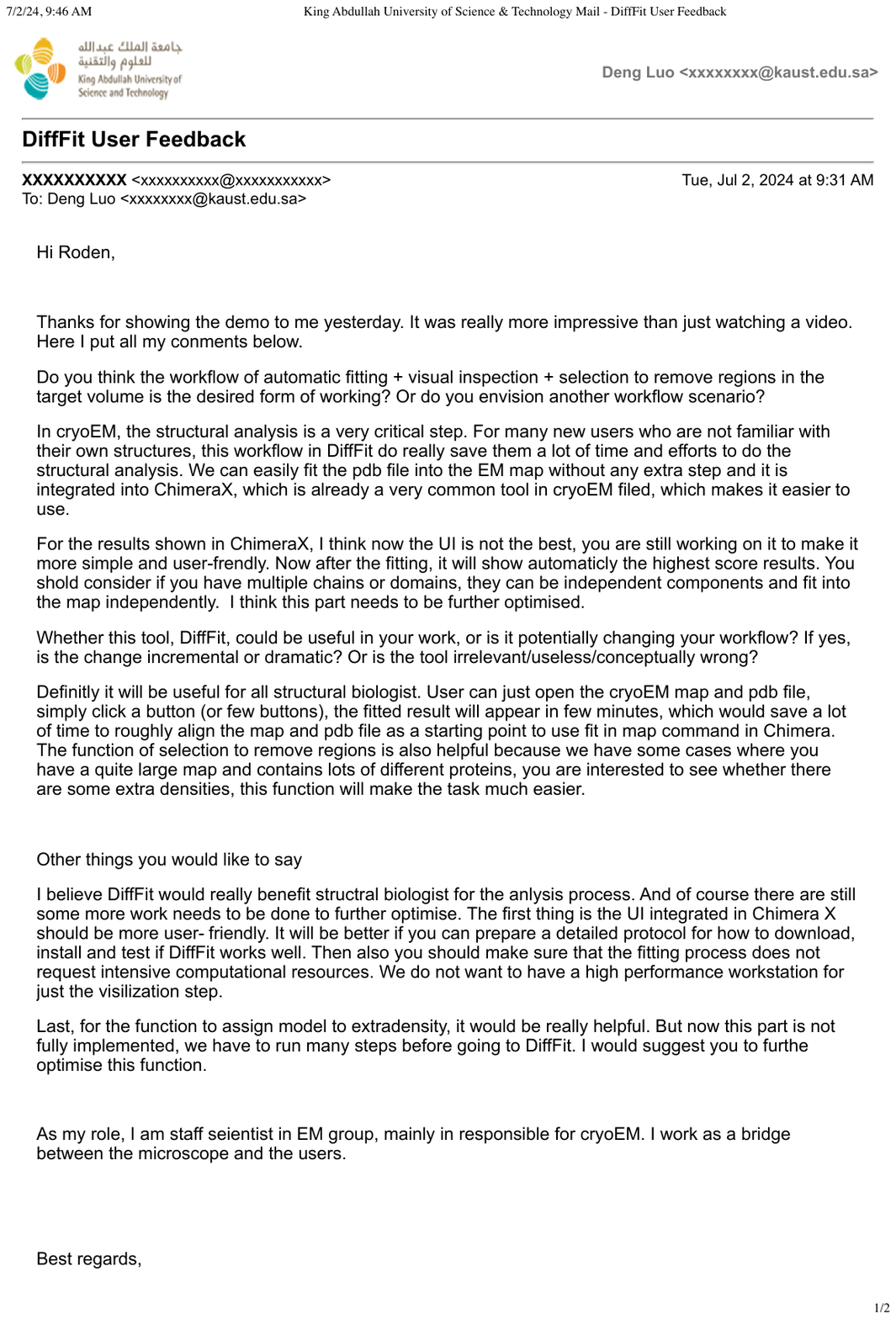}

\end{document}